\documentclass [12pt,a4paper]{article}
\usepackage {amsmath}
\usepackage{eqnarray}
\usepackage{relsize}
\usepackage {graphicx}
\graphicspath{{images/}}
\usepackage {comment}
\usepackage{authblk}
\usepackage{cite}
\usepackage{float}
\usepackage{cite}
\restylefloat{table}
\usepackage{multirow}  %unneeded    
\date{}
\title{Novel weak form quadrature elements for non-classical higher order beam and plate theories}
\author{Md Ishaquddin\thanks{Corresponding author: \textit{E-mail address: ishaquddinmd@iisc.ac.in}}, S.Gopalakrishnan\thanks {\textit{E-mail address: krishnan@iisc.ac.in; Phone: +91-80-22932048}}}

\begin{document}

\maketitle
\vspace{-10mm}
\noindent\textit{Department of Aerospace Engineering, Indian Institute of Science 
Bengaluru 560012, India}\\

\begin{abstract}

Based on Lagrange and Hermite interpolation two novel versions of weak form quadrature element are proposed for a non-classical Euler-Bernoulli beam theory. By extending these concept two new plate elements are formulated using Lagrange-Lagrange and mixed Lagrange-Hermite interpolations for a non-classical Kirchhoff plate theory. The non-classical theories are governed by sixth order partial differential equation and have deflection, slope and curvature as degrees of freedom. A novel and generalize way is proposed herein to implement these degrees of freedom in a simple and efficient manner. A new procedure to compute the modified weighting coefficient matrices for beam and plate elements is presented. The proposed elements have displacement as the only degree of freedom in the element domain and displacement, slope and curvature at the boundaries. The Gauss-Lobatto-Legender quadrature points are considered as element nodes and also used for numerical integration of the element matrices. The framework for computing the stiffness matrices at the integration points is analogous to the conventional finite element method. Numerical examples on free vibration analysis of gradient beams and plates are presented to demonstrate the efficiency and accuracy of the proposed elements.\\ 

\end{abstract}
\section*{1.0 INTRODUCTION}

In recent decades the research in the field of computational solid and fluid mechanics focused on developing cost effective and highly accurate numerical schemes. Subsequently, many numerical schemes were proposed and applied to various engineering problems. The early research emphasized on the development of finite element and finite difference methods\cite{zinki1,zinki2,bhate}, these methodologies had limitations related to the computational cost. Alternatively, differential quadrature method (DQM) was proposed by Bellman \cite{Bellman} which employed less number of grid points. Later, many enriched versions of differential quadrature method were developed, for example, differential quadrature method \cite{Bert1,Bert2,Shu,Du1,Du2,Wangb}, harmonic differential quadrature method\cite{Civalek1,Civalek2}, strong form differential quadrature element method (DQEM) \cite{Wang1,Wang2,Wang3,Xing,Karami1,Karami2,Karami3}, and weak form quadrature element method \cite{Striz1,Striz2,Zhong,Wang4}. The main theme in these improved DQ versions was to develop versatile models to account for complex loading, discontinuous geometries and generalized boundary conditions. \\
\indent Lately, much research inclination is seen towards the strong and weak form DQ methods due their versatality\cite{Wang1,Wang2,Wang3,Xing,Karami1,Karami2,Karami3,Striz1,Striz2,Zhong,Wang4}. The strong form differential quadrature method which is built on governing equations, require explicit expressions for interpolation functions and their derivatives, and yield unsymmetric matrices. In contrast, the weak form quadrature method is fomulated using variation principles, and the weighting coefficients are computed explicitly at integration points using the DQ rule, leading to symmetric matrices. The aforementioned literature forcussed on developing DQ schemes for classcial beam and plate theories which are governed by fourth order partial differential equations. The DQ solution for the sixth and eighth order differential equations using GDQR technique is due to Wu et al. \cite{Six,Eigth}. In their research, they employed strong form of governing equation in conjunction with Hermite interpolation function to compute the weighting coefficients and demonstrated the capability for structural and fluid mechanics problems. Recently, Wang et al. \cite{Wangnew} proposed a strong form differential quadrature element based on Hermite interpolation to solve a sixth order partial differential equation associated with a non-local Euler-Bernoulli beam. The capability of the element was demonstrated through free vibration analysis. In this article the main focus is to propose a weak form quadrature beam and plate element for non-classical higher order theories, which are characterized by sixth order partial differential equations. As per the authors knowledge no such work is reported in the literature till date.  

The non-classical higher order theories unlike classical continuum theories are governed by sixth order partial differential equations\cite{Mindlin1,Mindlin2,Mindlin3,Fleck,Koiter,Aifantis1}. These non-classical continuum theories are modified versions of classical continuum theories incorporating higher order gradient terms in the constitutive relations. The higher order terms consists of stress and strain gradients accompanied with intrinsic length which accounts for micro and nano scale effects\cite{Mindlin1,Mindlin2,Mindlin3,Fleck,Koiter,Aifantis1}. These scale dependent non-classical theories are efficient in capturing the micro and nano scale behaviours of  structural systems\cite{Mindlin3,Fleck,Koiter}. One such class of non-classical gradient elasticity theory is the simplified theory by Mindlin et al. \cite{Mindlin3}, with one gradient elastic modulus and two classical lame$^{'}$ constant for structural applications \cite{Aifantis1,Aifantis2,Aifantis3}. This simlified theory was applied earlier to study the static, dynamic and buckling behaviour of gradient elastic beams \cite{Besko1b,Besko2b,Besko3b} and plates \cite{Besko1p,Besko2p,Besko3p} by developing analytical solutions. Pegios et al. \cite{Pegios} developed a  finite element model for static and stability analysis of gradient beams. The numerical solution of 2-D and 3-D gradient elastic structural problems using finite element and boundary element methods can be found in \cite{BEM}.  

In this paper, we propose for the first time, two novel versions of weak form quadrature beam elements to solve a sixth order partial differential equation encountered in higher order non-classical elasticity theories. The two versions of quadrature beam element are based on Lagrange and $C^{2}$ continuous Hermite interpolations, respectively. Further, we extend this concept and develop two new types of quadrature plate elements for gradient elastic plate theories. The first element employs Lagrange interpolation in $x$ and $y$ direction and second element is based on Lagrange-Hermite mixed interpolation with Lagrange interpolation in $x$ and Hermite in $y$ direction. These elements are formulated with the aid of variation principles, differential quadrature rule and Gauss Lobatto Legendre (GLL) quadrature rule. Here, the GLL points are used as element nodes and also to perform numerical integration to evaluate the stiffness and consistent mass matrices. The proposed elements have displacement, slope and curvature as the degrees of freedom at the element boundaries and only displacement in the domain. A new way to incorporate the non-classical boundary conditions associated with the gradient elastic beam and plate theory is proposed and implemented. The novelty in the proposed scheme is the way the classical and non-classical boundary conditions are represented accurately and with ease.  It should be noted that the higher order degrees of freedom at the boundaries are built into the formulation only to enforce the boundary conditions. 

The paper is organized as follows, first the theoretical basis of gradient elasticity theory required to formulate the quadrature elements is presented. Next, the quadrature elements based on Lagrange and Hermite interpolations functions for an Euler-Bernoulli gradient beam are formulated. Later, the formulation for the quadrature plate elements are given. Finally, numerical results on free vibration analysis of gradient beams and plates are presented to demonstrate the capability of the proposed elements followed by conclusions. 
	
\section{Strain gradient elasticity theory}
In this study, we consider Mindlin's \cite{Mindlin3} simplified strain gradient micro-elasticity theory with two classical and one non-classical material constants. The two classical material constants are Lame$^{'}$ constants and the non-classical one is related to intrinsic bulk length $g$. The theoretical basis of gradient elastic theory required to formulate the quadrature beam and plate elements are presented in this section.

\subsection{Gradient elastic beam theory} \label{section_Sg_Beam}

The stress-strain relation for a 1-D gradient elastic theory is given as \cite{Vardo,Besko1b}
\begin{align*}
{\tau}&= 2\,\,\mu \,\, \varepsilon + \lambda \,\, {\text{tr}} \varepsilon \,\, \text{I} \\
{{\varsigma}}&= g^{2} \,\, [2 \,\, \mu \,\, \nabla\varepsilon + \lambda \,\, \nabla (\text{tr}\varepsilon) \,\, \text{I}] \tag{1}
\end{align*}

\noindent where $\lambda$,$\,$ $\mu$ are Lam$e^{'}$ constants.$\nabla=\frac{\partial}{\partial x}+\frac{\partial}{\partial y}$ is the Laplacian operator and $\text{I}$ is the unit tensor. $\tau$, $\varsigma$ denotes Cauchy and higher order stress respectively, $\varepsilon$ and ($\text{tr}\,\varepsilon$) are the classical strain and its trace which are expressed in terms of displacement vector $\textit{w}$ as
\begin{align*}
&{\varepsilon}= \frac{1}{2}(\nabla\textit{w}+\textit{w}\nabla)\,\,, \,\,\quad \text{tr}{\varepsilon}= \nabla\textit{w} \tag{2}
\end{align*}

From the above equations the constitutive relations for an Euler-Bernoulli gradient beam can be defined as
\begin{align*}         
{\tau_{x}}= E\varepsilon_{x}, \quad \varsigma_{x}={g}^{2}\varepsilon_{x}^{'}, \quad \varepsilon_{x}=-z\dfrac{\partial^{2} w(x,t)}{\partial{x}^2}\tag{3}
\end{align*}

\noindent For the above state of stress and strain the strain energy in terms of displacements for a beam defined over a domain $-L/2\leq x \leq L/2$ can be written as \cite{Vardo}
\begin{align*}         
{U}= \frac{1}{2}\int_{-L/2}^{L/2} EI\big[(w^{''})^{2}+g^{2}(w^{'''})^{2}\big]dx  \tag{4}
\end{align*}
The kinetic energy is given as
\begin{align*}         
{K}= \frac{1}{2}\int_{t_{0}}^{t_{1}}\int_{-L/2}^{L/2}\rho{A}\dot{w}^{2}{dx}{dt}  \tag{5}
\end{align*}

\noindent  where $E$, $A$, $I$ and $\rho$ are the Young's modulus, area, moment of inertia, and density, respectively. $w(x,t)$ is transverse displacement and over dot indicates differentiation with respect to time.\\
 
Using the The Hamilton's principle\cite{Reddyb}:
\begin{align*}         
\delta\int_{t_{0}}^{t_{1}}(U-K)\,dt=0   \tag{6}
\end{align*}

\noindent we get the following weak form expression for elastic stiffness matrix `K' and consistent mass matrix `m' as 
\begin{align*}   \label{K_beam}      
&K= \int_{-L/2}^{L/2} EI\big[w{''}\,\delta{w{''}}+g^{2}	\,w^{'''}\delta{w}^{'''}\big]dx \tag{7}
\end{align*}
\begin{align*}   \label{M_beam}      
 &m=\int_{-L/2}^{L/2}\,\rho{A}\,\dot{w}\,\dot{\delta{w}}\,dx \tag{8}
\end{align*}

\noindent The governing equation of motion for a gradient elastic Euler-Bernoulli beam is obtained as
\begin{align*}   \label{EOM_beam}      
EI(w^{\textit{iv}}-{g}^{2}w^{\textit{vi}})+m\ddot{w}=0 \tag{9}
\end{align*}

The above sixth order equation of motion yields three independent variables related to deflection $w$, slope $w^{'}$ and curvature $w^{''}$ and six boundary conditions in total, as given below\\

\noindent \textit{Classical boundary conditions} :     
\begin{align*}       
V&=EI[w^{'''}-{g}^{2}w^{v}]=0\hspace{0.3cm}\text{or}\,\, w=0,\hspace{0.3cm}\text{at}\,\,x=(-L/2,L/2) \\ \tag{10}
M&=EI[w^{''}-{g}^{2}w^{iv}]=0\hspace{0.3cm}\text{or}\,\, w^{'}=0,\hspace{0.3cm}\text{at}\,\,x=(-L/2,L/2)\\  
\end{align*}
\noindent \textit { Non-classical boundary conditions} :         
\begin{align*}         
\bar{M}&=[{g}^{2}EIw^{'''}]=0 \hspace{0.3cm}\text{or}\,\,\, w^{''}=0,\hspace{0.3cm}\text{at}\,\,x=(-L/2,L/2) 		\tag{11}
\end{align*}

where $V$, $M$ and $\bar{M}$ are shear force, bending moment and higher order moment, respectively.

\subsection{Gradient elastic plate theory} \label{section_Sg_Plate}
 
The strain-displacement relations for a Kirchhoff's plate theory are defined as \cite{Timo}
\begin{align*}
{\varepsilon_{xx}}= -z{\bar{w}}_{xx}, \quad {\varepsilon_{yy}}=-z{\bar{w}}_{yy} , \quad \gamma_{xy}=-2z{\bar{w}}_{xy} \tag{12}
\end{align*}

\noindent where $\bar{w}(x,y,t)$  is transverse displacement of the plate. 
The stress-strain relations for a gradient elastic Kirchhoff plate are given by \cite{Vardo,Koiter}:\\

\noindent \textit{Classical:}
\begin{align*}
{\tau}_{xx}= &\frac{E}{1-\nu^{2}}(\varepsilon_{xx}+\nu\varepsilon_{yy})\\
{\tau}_{yy}= &\frac{E}{1-\nu^{2}}(\varepsilon_{yy}+\nu\varepsilon_{xx})\tag{13} \\
{\tau}_{xy}= &\frac{E}{1+\nu}\varepsilon_{xy}  
\end{align*}
\noindent \textit{Non-classical:}
\begin{align*}
{\varsigma_{xx}}= &g^{2} \frac{E}{1-\nu^{2}} \nabla^{2}(\varepsilon_{xx}+\nu\varepsilon_{yy})\\
{\varsigma_{yy}}= &g^{2} \frac{E}{1-\nu^{2}}\nabla^{2}(\varepsilon_{yy}+\nu\varepsilon_{xx})\tag{14} \\
{\varsigma_{xy}}= &g^{2} \frac{E}{1+\nu}\nabla^{2}\varepsilon_{xy} 
\end{align*}

\noindent where $\tau_{xx}$, $\tau_{yy}$,$\tau_{xy}$, are the classical Cauchy stresses and $\varsigma_{xx}$,$\varsigma_{yy}$, $\varsigma_{xy}$ denotes higher order stresses related to gradient elasticity. The strain energy for a gradient elastic Kirchhoff plate is gven by \cite{Koiter,Besko3p}
\begin{align*}        
						U_{p}=U_{cl}+U_{sg} \tag{15}
\end{align*}

\noindent where $U_{cl}$ and $U_{sg}$ are the classical and gradient elastic strain energy given by
\begin{align*}        
{U}_{cl}= \frac{1}{2}D\,\int\int_{A}\Big[\bar{w}_{xx}^{2}+\bar{w}_{yy}^{2}+2\bar{w}_{xy}^{2}+2\,\nu\,(\bar{w}_{xx}\bar{w}_{yy}
-\bar{w}_{xy}^{2})\Big]dxdy \tag{16}
\end{align*}
\begin{align*}        
{U}_{sg}= &\frac{1}{2}g^{2}D\,\int\int_{A}\Big[\bar{w}_{xxx}^{2}+\bar{w}_{yyy}^{2}+3(\bar{w}_{xyy}^{2}+
\bar{w}_{xxy}^{2})\\& \hspace{2.0cm}\,\,\,\,+ 2\,\nu\,(\bar{w}_{xyy}\bar{w}_{xxx}+ \bar{w}_{xxy}\bar{w}_{yyy}-\bar{w}_{xyy}^{2}-\bar{w}_{xxy}^{2}\Big]dxdy \tag{17}
\end{align*}
 	 	
\noindent where, $D=\frac{E{h}^3}{12(1-\nu^{2})}$.\\
 	 	
The kinetic energy is given by
\begin{align*}         
{K}= \frac{1}{2}\int_{A}\,\rho\,{h}\,\dot{\bar{w}}^{2}\,{dx}\,{dy}\ \tag{18}
\end{align*}

Using the The Hamilton's principle:
\begin{align*}         
\delta\int_{t_{0}}^{t_{1}}(U-K)\,dt=0   \tag{19}
\end{align*}

\noindent we obtain the following expression for elastic stiffness and mass matrix for a gradient elastic plate\\

$Elastic \,stiffness\, matrix$ :
\begin{align*}       \label{K_plate}  
[K]=[K]_{cl}+[K]_{sg} \tag{20}
\end{align*}

\noindent where $[K]_{cl}$, $[K]_{sg}$ are classical and non-classical elastic stiffness matrix defined as
\begin{align*}         
[K]_{cl}= &D\int_{A}\Big[\bar{w}_{xx}\,\delta{\bar{w}}_{xx}+\bar{w}_{yy}\,\delta{\bar{w}}_{yy}+2\bar{w}_{xy}\,\delta{\bar{w}}_{xy}+\,\\ & \hspace{2.0cm} \nu\,(\delta{\bar{w}}_{xx}\,\bar{w}_{yy}+\bar{w}_{xx}\,\delta{\bar{w}}_{yy}
-2\bar{w}_{xy}\,\delta{\bar{w}}_{xy})\Big]dxdy \tag{21}
\end{align*}
\begin{align*}        
[K]_{sg}= &g^{2}D\,\int_{A}\Big[\bar{w}_{xxx}\,\delta{\bar{w}}_{xxx}+\bar{w}_{yyy}\,\delta{\bar{w}}_{yyy}+3(\bar{w}_{xyy}\,\delta{\bar{w}}_{xyy}+\\& \hspace{1.0cm}\,\,\,\,
\bar{w}_{xxy}\,\delta{\bar{w}}_{xxy})+ \nu\,(\bar{w}_{xyy}\,\delta{\bar{w}}_{xxx}+\bar{w}_{xxx}\,\delta{\bar{w}}_{xyy}+\\& \hspace{1.0cm}\,\,\,\, \bar{w}_{xxy}\,\delta{\bar{w}}_{yyy}+\bar{w}_{yyy}\,\delta{\bar{w}}_{xxy}-2\,\bar{w}_{xyy}\,\delta{\bar{w}}_{xyy}-2\,\bar{w}_{xxy}\,\delta{\bar{w}}_{yxx})\Big]dxdy \tag{22}
\end{align*}

$Consistent\, mass\, matrix$ :
\begin{align*}    \label{M_plate}     
[M]= \int_{A}\,\rho\,{h}\,\dot{\bar{w}}\,\delta{\dot{\bar{w}}}\,{dx}\,{dy}\ \tag{23}
\end{align*}

The equation of motion for a gradient elastic Kirchhoff plate considering the inertial effect is obtained as:
\begin{align*}    \label{EOM_plate}     
D\nabla^{4}\bar{w}-g^{2}D\nabla^{6}\bar{w}+\rho h\frac{\partial^{2}\bar{w}}{\partial t^{2}}=0\tag{24} 
\end{align*}
where,
\begin{align*}         
&\nabla^{4}=\frac{\partial^{4} \bar{w}}{\partial x^{4}}+\frac{\partial^{4} \bar{w}}{\partial y^{4}}+2\frac{\partial^{4} \bar{w}}{\partial x^{2}\partial y^{2}},\\ 
&\nabla^{6}=\frac{\partial^{6} \bar{w}}{\partial x^{6}}+\frac{\partial^{6} \bar{w}}{\partial y^{6}}+3\frac{\partial^{6} \bar{w}}{\partial x^{4}\partial y^{2}}+3\frac{\partial^{6} \bar{w}}{\partial x^{2}\partial y^{4}}
\end{align*}

\noindent the associated boundary conditions for the plate with origin at $(0,0)$ and domain defined over ($-l_{x}/2\leq x \leq l_{x}/2$), ($-l_{y}/2\leq y \leq l_{y}/2$), are listed below.\\ 

\noindent \textit{Classical boundary conditions} :\\
\begin{align*}    
 V_{x}=-D\bigg(\frac{\partial^{3} \bar{w}}{\partial x^{3}}+(2-\nu) \frac{\partial^{3} \bar{w}}{\partial x\partial y^{2}}\bigg)+g^{2}D\bigg[ \frac{\partial^{5} \bar{w}}{\partial x^{5}}+(3-\nu) \frac{\partial^{5} \bar{w}}{\partial x\partial y^{4}}+3 \frac{\partial^{5} \bar{w}}{\partial y^{2}\partial x^{3}}\bigg]=0 
 \\ \text{or}\\ \,\, \bar{w}=0\,\, \text{at}\,\,x=(-l_{x}/2,l_{x}/2)
\end{align*}
\begin{align*}    
V_{y}=-D\bigg(\frac{\partial^{3} \bar{w}}{\partial y^{3}}+(2-\nu) \frac{\partial^{3} \bar{w}}{\partial y\partial x^{2}}\bigg)+g^{2}D\bigg[ \frac{\partial^{5} \bar{w}}{\partial y^{5}}+(3-\nu) \frac{\partial^{5} \bar{w}}{\partial y\partial x^{4}}+3 \frac{\partial^{5} \bar{w}}{\partial x^{2}\partial y^{3}}\bigg]=0 \\ \text{or}\\ \,\, \bar{w}=0,
\text{at}\,\,y=(-l_{y}/2,l_{y}/2) \\ \tag{25}
\end{align*}
\begin{align*}    
M_{x}=-D\bigg(\frac{\partial^{2} \bar{w}}{\partial x^{2}}+\nu \frac{\partial^{2} \bar{w}}{\partial y^{2}}\bigg)+g^{2}D\bigg[ \frac{\partial^{4} \bar{w}}{\partial x^{4}}+\nu \frac{\partial^{4} \bar{w}}{\partial y^{4}}+(3-\nu) \frac{\partial^{4} \bar{w}}{\partial x^{2}\partial y^{2}}\bigg]=0  &\\ \text{or} \\ \, \bar{w}_{x}=0,\hspace{0.1cm}\text{at}\,\,x=(-l_{x}/2,l_{x}/2)\\  \\  
M_{y}=-D\bigg(\frac{\partial^{2} \bar{w}}{\partial y^{2}}+\nu \frac{\partial^{2} \bar{w}}{\partial x^{2}}\bigg)+g^{2}D\bigg[ \frac{\partial^{4} \bar{w}}{\partial y^{4}}+\nu \frac{\partial^{4} \bar{w}}{\partial x^{4}}+(3-\nu) \frac{\partial^{4} \bar{w}}{\partial x^{2}\partial y^{2}}\bigg]=0 &\\ \text{or} \\ \, \bar{w}_{y}=0,\hspace{0.1cm}\text{at}\,\,y=(-l_{y}/2,l_{y}/2)\\  
\tag{26}
\end{align*}
\noindent \textit { Non-classical boundary conditions} :  
\begin{align*}         
\bar{M}_{x}=-g^{2}D\bigg(\frac{\partial^{3} \bar{w}}{\partial x^{3}}+\nu \frac{\partial^{3} \bar{w}}{\partial x\partial y^{2}}\bigg)=0 \,\,\text{or}\,\,\, \bar{w}_{xx}=0,\,\,\,\text{at}\,\,x=(-l_{x}/2,l_{x}/2)	\\	
\bar{M}_{y}=-g^{2}D\bigg(\frac{\partial^{3} \bar{w}}{\partial y^{3}}+\nu \frac{\partial^{3} \bar{w}}{\partial y\partial x^{2}}\bigg)=0 \,\,\text{or}\,\,\, \bar{w}_{yy}=0,\,\,\,\text{at}\,\,y=(-l_{y}/2,l_{y}/2) \\  \tag{27}
\end{align*}

\noindent  where $l_{x}$ and $l_{y}$ are the length and width of the plate. $V_{x}$,$V_{y}$ are the shear force, $M_{x}$,$M_{y}$ are the bending moment and $\bar{M}_{x}$,$\bar{M}_{y}$ are the higher order moment. The different boundary conditions employed in the present study for a gradient elastic Kirchhoff plate are\\

\noindent \textit{Simply supported on all edges SSSS}:\\
\noindent $\bar{w}=M_{x}=\bar{w}_{xx}=0$ \hspace{0.2cm} at\hspace{0.3cm} $x=(-l_{x}/2,l_{x}/2)$ \\
\noindent $\bar{w}=M_{y}=\bar{w}_{yy}=0$ \hspace{0.2cm} at\hspace{0.3cm} $y=(-l_{y}/2,l_{y}/2)$ \\

\noindent \textit{Free on all edges FFFF}:\\
\noindent $V_{x}=M_{x}=\bar{M}_{x}=0 $  \hspace{0.2cm} at \hspace{0.3cm}$x=(-l_{x}/2,l_{x}/2)$ \\
\noindent $V_{y}=M_{y}=\bar{M}_{y}=0 $  \hspace{0.2cm} at \hspace{0.3cm}$y=(-l_{y}/2,l_{y}/2)$ \\

\noindent \textit{Simply supported and free on adjacent edges SSFF}:\\
\noindent $\bar{w}=M_{y}=\bar{w}_{yy}=0$ \hspace{0.25cm} at \hspace{0.3cm}$y=-l_{y}/2$ \\
\noindent $\bar{w}=M_{x}=\bar{w}_{xx}=0$ \hspace{0.2cm} at \hspace{0.3cm}$x=l_{x}/2$ \\
\noindent $V_{x}=M_{x}=\bar{M}_{x}=0 $ \hspace{0.15cm} at \hspace{0.3cm}$x=-l_{x}/2$ \\
\noindent $V_{y}=M_{y}=\bar{M}_{y}=0 $ \hspace{0.15cm} at \hspace{0.3cm}$y=l_{y}/2$ \\

\noindent for the SSFF plate at $(-l_x/2,-l_y/2)$ and $(l_x/2,l_y/2)$, $\bar{w}=0$ condition is enforced. The above boundary conditions are described by a notation, for example, consider a SSFF plate, the first and second letter correspond to $y=-l_{y}/2$ and $x=l_{x}/2$ edges, similarly, the third and fourth letter correspond to the edges $y=l_{y}/2$ and $x=-l_{x}/2$, respectively. Further, the letter S, C and F correspond to simply supported, clamped and free edges of the plate.

\section{Quadrature element for a gradient elastic Euler-Bernoulli beam} \label{beam_qem}

\begin{figure}[H]
\includegraphics[width=1.0\textwidth]{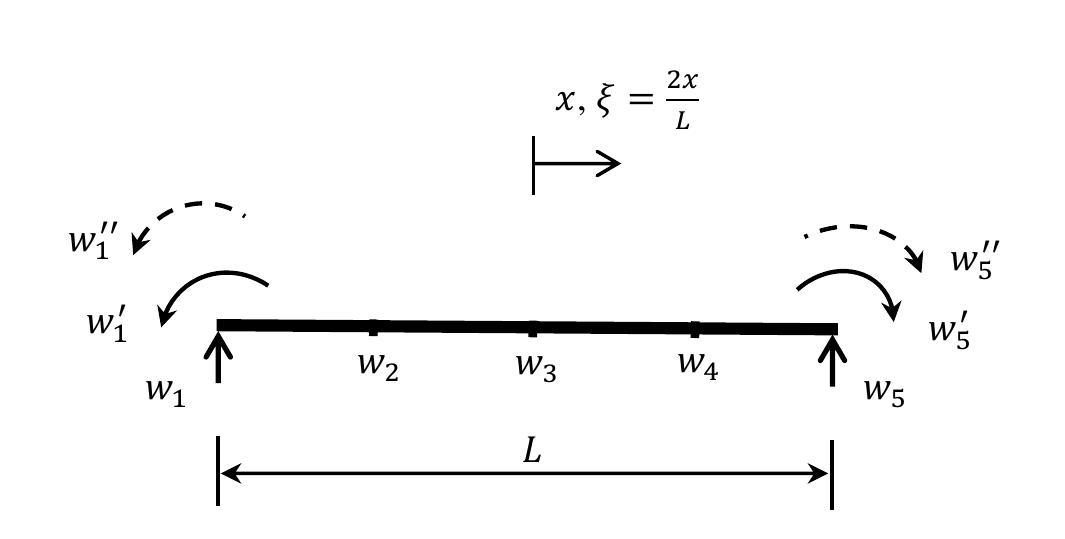}
\centering
\caption{A typical quadrature element for a gradient Euler-Bernoulli beam.}
\label{fig:beam}
\end{figure}

Two novel quadrature elements for a gradient Euler-Bernoulli beam are presented in this section. First, the quadrature element based on Lagrangian interpolation is formulated. Later, the quadrature element based on $C^{2}$ continuous Hermite interpolation is developed. The procedure to modify the DQ rule to implement the classical and non-classical boundary conditions are explained.  A typical N-node quadrature element for an Euler-Bernoulli gradient beam is shown in the Figure \ref{fig:beam}. 
It can be noticed from the Figure \ref{fig:beam}, each interior node has only displacement $w$ as degrees of freedom and the boundary has 3 degrees of freedom $w$, $w^{'}$, $w^{''}$. The new displacement vector now includes the slope and curvature as additional degrees of freedom at the element boundaries given by: $w^{b}=\{w_{1},\cdots, w_{N},w^{'}_{1},w^{'}_{N},w^{''}_{1},w^{''}_{N}\}$. The procedure to incorporate these extra boundary degrees of freedom in to the formulation will be presented next for Lagrange and $C^{2}$ continuous Hermite interpolation based quadrature elements.

\subsection{Lagrange interpolation based quadrature beam element} \label{section_Lag_qem_beam}
The displacement for a N-node quadrature beam is assumed as\cite{Wangb}:
\begin{align*}    \label{disp_Lag_qem_beam}
w(x,t)=\sum_{j=1}^{N} L_{j}(x)w^{b}_{j}=\sum_{j=1}^{N} \bar{L}_{j}(\xi)w^{b}_{j} \tag{28}
\end{align*}

\noindent $L_{j}(x)$ and $\bar{L}_{j}(\xi)$ are Lagrangian interpolation functions in $x$ and $\xi$ co-ordinates respectively, and $\xi=2x/L$ with $\xi \in[-1,1]$. The Lagrange interpolation functions are defined as\cite{Wangb,Shu}
\begin{align*}    
L_{j}(\xi)=\frac{\beta(\xi)}{\beta(\xi_{j})}=\prod_{\substack{k=1 \\ (k\neq j)}}^{N}\frac{(\xi-\xi_{k})}{(\xi_{j}-\xi_{k})}   \tag{29}
\end{align*}
\noindent where \\
 $\beta(\xi)=(\xi-\xi_{1})(\xi-\xi_{2})\cdots(\xi-\xi_{j-1})(\xi-\xi_{j+1})\cdots(\xi-\xi_{N})$ \\
 $\beta(\xi_{j})=(\xi_{j}-\xi_{1})(\xi_{j}-\xi_{2})\cdots(\xi_{j}-\xi_{j-1})(\xi_{j}-\xi_{j+1})\cdots)(\xi_{j}-\xi_{N})$ \\

\noindent The first order derivative of the above interpolation function can be written as,
\begin{align*}    
A_{ij}={L}^{'}_{j}(\xi_{i})\begin{cases}
\mathlarger\prod_{\substack{k=1 \\ (k\neq i,j)}}^{N}(\xi_{i}-\xi_{k})/\mathlarger\prod_{\substack{k=1 \\ (k\neq j)}}^{N}=(\xi_{j}-\xi_{k})\,\,\,\, (i\neq j)\\ \\ 
\mathlarger\sum_{\substack{k=1 \\ (k\neq i)}}^{N}\frac{1}{(\xi_{i}-\xi_{k})}
\end{cases}\tag{30}  
\end{align*}

The conventional higher order weighting coefficients are computed as
\begin{align*}    
B_{ij}=\sum_{k=1}^{N} A_{ik}A_{kj} \,\, , \quad
C_{ij}=\sum_{k=1}^{N} B_{ik}A_{kj}  \,\,\,(i,j=1,2,...,N)\tag{31}
\end{align*}\\
Here, $B_{ij}$ and $C_{ij}$ are weighting coefficients for second and third order derivatives, respectively.

The sixth order partial differential equation given in Equation (9) renders  slope $w^{'}$ and curvature $w^{''}$ as extra degrees of freedom at the element boundaries. To account for these extra boundary degrees of freedom in the formulation, the derivatives of conventional weighting function $A_{ij}$, $B_{ij}$, and $C_{ij}$ are modified as follows:\\ 

\noindent \textit{First order derivative matrix}: 
\begin{align*}    \label{Lag_Aij}
\bar{A}_{ij}=\begin{cases}
A_{ij} \,\,\,\, (i,j=1,2,\cdots,N)\\ \\
0 \,\,\,\,\,\,\,(i,j=1,2,\cdots,N,j=N+1,\cdots,N+4)\tag{32}
\end{cases} 
\end{align*}
\noindent \textit{Second order derivative matrix}: 
\begin{align*} \label{Lag_Bij1}
\bar{B}_{ij}=\begin{cases}
B_{ij} \,\,\,\, (j=1,2,\cdots,N)\\ \\
0 \,\,\,\,\,\,\,(j=N+1,\cdots,N+4)  (i=2,3,\cdots,N-1)\tag{33}
\end{cases} 
\end{align*}
\begin{align*} \label{Lag_Bij2}
\bar{B}_{ij}=\sum_{k=2}^{N-1}A_{ik}A_{kj} \,\,\, (j=1,2,\cdots,N,i=1,N)\\ 
\bar{B}_{i(N+1)}=A_{i1}\,\,; \,\, \,\,\,  \,\,\bar{B}_{i(N+2)}=A_{iN} \,\,\,\,(i=1,N)\tag{34}
\end{align*} 

\noindent \textit{Third order derivative matrix}: 
\begin{align*}\label{Lag_Cij1}
\bar{C}_{ij}=\begin{cases}
C_{ij} \,\,\,\, (j=1,2,\cdots,N)\\ \\
0 \,\,\,\,\,\,\,(j=N+1,\cdots,N+4, i=2,3,\cdots,N-1)\tag{35}
\end{cases}
\end{align*} 
\begin{align*} \label{Lag_Cij2}
\bar{C}_{ij}=\sum_{k=2}^{N-1}B_{ik}A_{kj}\,\,\  (j=1,2,\cdots,N,i=1,N)\\ 
\bar{C}_{i(N+3)}=A_{i1}\,\,; \,\, \,\, \,\,\, \bar{C}_{i(N+4)}=A_{iN} \,\,\,\,(i=1,N)\tag{36}
\end{align*} 

 Using the above Equations (\ref{Lag_Aij})-(\ref{Lag_Cij2}), the element matrices can be expressed in terms of weighting coefficients as \\
	 
\noindent $Elastic \,stiffness\, matrix$ :
\begin{align*}     \label{Lag_Kij_dqm_beam}    
K_{ij}=\frac{8EI}{L^{3}}\sum_{k=1}^{N} {H}_{k}\bar{B}_{ki}\bar{B}_{kj}+g^{2}\frac{32EI}{L^{5}}\sum_{k=1}^{N} {H}_{k}\bar{C}_{ki}\bar{C}_{kj} \\ \,\,\,\,\,
(i,j=1,2,...,N,N+1,\cdots,N+4)\tag{37}
\end{align*}

\noindent $Consistent\, mass\, matrix$ :
\begin{align*}     \label{Lag_Mij_dqm_beam}     
M_{ij}=\frac{\rho{A}L}{2}{H}_{k}\delta_{ij}\,\,\,\,\,
(i,j=1,2,...,N)\tag{38}
\end{align*}

Here $\xi$ and $H$ are the coordinate and weights of GLL quadrature.      
 $\delta_{ij}$ is the Dirac-delta function. 

\subsection{Hermite interpolation based quadrature beam element} \label{section_Herm_qem_beam}

For the case of quadrature element based on $C^{2}$ continuous Hermite interpolation the displacement for a N-node gradient beam element is assumed as 
\begin{align*}    \label{disp_Herm_qem_beam}
w(\xi,t)=\sum_{j=1}^{N} \phi_{j}(\xi)w_{j}+\psi_{1}(\xi)w_{1}^{'}+\psi_{N}(\xi)w_{N}^{'}+\varphi_{1}(\xi)w_{1}^{''}+\varphi_{N}(\xi)w_{N}^{''}=\sum_{j=1}^{N+4} \Gamma_{j}(\xi)w^{b}_{j}\tag{39}
\end{align*}

\noindent $\phi$, $\psi$ and $\varphi$ are Hermite interpolation functions defined as \cite{Six,Wangnew}
\begin{align*}    \label{Herm_Aij1}
\varphi_{j}(\xi)=\frac{1}{2(\xi_{j}-\xi_{N-j+1})^{2}}L_{j}(x)(x-x_{j})^{2}(x-x_{N-j+1})^{2}  (j=1, N)\tag{40}
\end{align*}
\begin{align*}    \label{Herm_Bij1}
\psi_{j}(\xi)=\frac{1}{(\xi_{j}-\xi_{N-j+1})^{2}}L_{j}(\xi)(\xi-\xi_{j})(\xi-\xi_{N-j+1})^{2}\\
-\bigg[2L_{j}^{1}(\xi
_{j})+\frac{4}{\xi_{j}-\xi_{N-j+1}}\bigg]\varphi
_{j}(\xi)  \,\,\, (j=1, N)\tag{41}
\end{align*}
\begin{align*}    \label{Herm_Cij1}
\phi_{j}(\xi)=\frac{1}{(\xi_{j}-\xi_{N-j+1})^{2}}L_{j}(\xi)(\xi-\xi_{N-j+1})^{2} -\bigg[L_{j}^{1}(\xi
_{j})+\frac{2}{\xi_{j}-\xi_{N-j+1}}\bigg]\psi 
_{j}(\xi)\\-\bigg[L_{j}^{2}(\xi
_{j})+\frac{4L_{j}^{1}(\xi
_{j})}{\xi_{j}-\xi_{N-j+1}}+\frac{2}{(\xi_{j}-\xi_{N-j+1})^{2}}\bigg]\varphi
_{j}(\xi)  \,\,\, (j=1, N)\tag{42}
\end{align*}
\begin{align*}    \label{Herm_Dij1}
\phi_{j}(\xi)=\frac{1}{(\xi_{j}-\xi_{1})^{2}(\xi_{j}-\xi_{N})^{2}}L_{j}(\xi)(\xi-\xi_{1})^{2}(\xi-\xi_{N})^{2}   \,\,\, (j=2,3,...,N-1)\tag{43}
\end{align*}

The $k$th order derivative of $w(\xi)$ with respect to $\xi$ is obtained from Equation (\ref{disp_Herm_qem_beam}) as 
\begin{align*}   \label{Herm_Aij_main} 
w^{k}(\xi)=\sum_{j=1}^{N} \phi_{j}^{k}(\xi)w_{j}+\psi_{1}^{k}(\xi)w_{1}^{'}+\psi_{N}^{k}(\xi)w_{N}^{'}+\varphi_{1}^{k}(\xi)w_{1}^{''}+\varphi_{N}^{k}(\xi)w_{N}^{''}=\sum_{j=1}^{N+4} \Gamma_{j}^{k}(\xi)w^{b}_{j}\tag{44}
\end{align*}

 Using the above Equation (\ref{Herm_Aij1})-(\ref{Herm_Aij_main}), the element matrices can be expressed in terms of weighting coefficients as \\

\noindent $Elastic \,stiffness\, matrix$ :
\begin{align*}    \label{Herm_Kij_dqm_beam}
K_{ij}=\frac{8EI}{L^{3}}\sum_{k=1}^{N} {H}_{k}\Gamma_{ki}^{(2)}\Gamma_{kj}^{(2)}+g^{2}\frac{32EI}{L^{5}}\sum_{k=1}^{N} {H}_{k}\Gamma_{ki}^{(3)}\Gamma_{kj}^{(3)} \\ \,\,\,\,\,
(i,j=1,2,...,N,N+1,\cdots,N+4)\tag{45}
\end{align*}
\noindent here $\xi$ and $H$ are the coordinate and weights of GLL quadrature. The consistent mass matrix remains the same as given by Equation (\ref{Lag_Mij_dqm_beam}).

Combining the stiffness and mass matrix, the system of equations after applying the boundary conditions can be expressed as
\begin{align*}\label{eq:Boundary_disp_Beam}
\begin{bmatrix}
\,k_{bb} & \phantom{-}k_{bd} \\ \\ 
\,k_{db} & \phantom{-}k_{dd} \\ \\ 
 \end{bmatrix}\begin{Bmatrix}
\,\Delta_{b} \\ \\ 
\,\Delta_{d}  \\ \\ 
 \end{Bmatrix}=\begin{bmatrix}
\,I & \phantom{-}0 \\ \\ 
\,0 & \phantom{-}\omega^{2}M_{dd} \\ \\ 
 \end{bmatrix}\begin{Bmatrix}
\,f_{b} \\ \\ 
\,\Delta_{d}  \\ \\ 
 \end{Bmatrix} \tag{46}
\end{align*}

\noindent where the vector $\Delta_{b}$ contains the boundary related non-zero slope and curvature dofs. Similarly, the vector $\Delta_{d}$ includes all the non-zero displacement dofs of the beam. In the present analysis the boundary force vector is assumed to be zero, $f_{b}=0$. Now, expressing the $\Delta_{b}$ dofs in terms of $\Delta_{d}$, the system of equations reduces to 
\begin{align*} \label{eq:Domain_disp_Beam}
\Big[k_{dd}-k_{db}k_{bb}^{-1}k_{bd}\Big] {\Big\{}\Delta_{d}{\Big\}}=
\omega^{2}\Big[M_{dd}\Big]{\Big\{}w_{d}{\Big\}}\tag{47}
\end{align*}

Here, $\bar{K}=\Big[k_{dd}-k_{db}k_{bb}^{-1}k_{bd}\Big]$ is the modified stiffness matrix associated with $\Delta_{d}$ dofs. The above system of equations leads to an Eigenvalue problem and its solutions renders frequencies and corresponding mode shapes.

\section{Quadrature element for gradient elastic Kirchhoff plate}

\begin{figure}[H]
\includegraphics[width=1.0\textwidth]{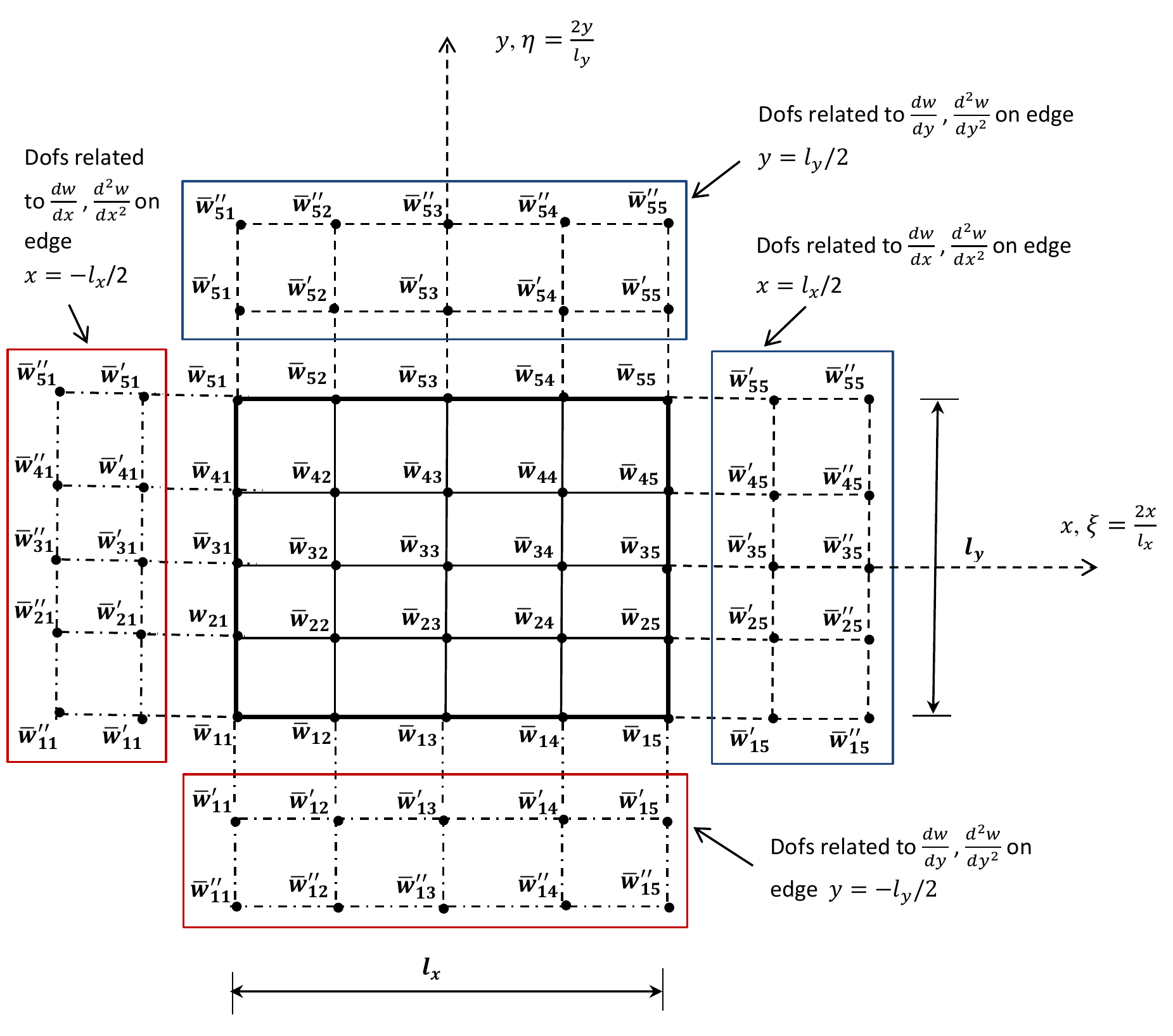}
\centering
\caption{A typical quadrature element for a gradient elastic Kirchhoff plate with $N=N_{x}=N_{y}=5$.}
\label{fig:Plate}
\end{figure}

In this section, we formulate two novel quadrature elements for non-classical gradient Kirchhoff plate. First, the quadrature element based on Lagrange interpolation in $\xi$ and $\eta$ direction is presented. Next, the quadrature element based on Lagrange-Hermite mixed interpolation, with Lagrangian interpolation is $\xi$ direction and Hermite interpolation assumed in $\eta$ direction is formulated. The GLL points in $\xi$ and $\eta$ directions are used as element nodes. Similar to the beam elements discussed in the section \ref{beam_qem}, the plate element also has displacement $\bar{w}$ as the only degrees of freedom in the domain and at the edges it has 3 degrees of freedom $\bar{w}$, $\bar{w}_{x}$ or $\bar{w}_{y}$, $\bar{w}_{xx}$ or $\bar{w}_{yy}$ depending upon the edge. At the corners the element has five degrees of freedom, $\bar{w}$, $\bar{w}_{x}$, $\bar{w}_{y}$, $\bar{w}_{xx}$ and $\bar{w}_{yy}$. The new displacement vector now includes the slope and curvature as additional degrees of freedom at the element boundaries given by: $w^{p}=\{\bar{w}_{i},\cdots,\bar{w}_{N\times N},\bar{w}^{j}_{x},\cdots,\bar{w}^{j}_{y},\cdots,\bar{w}^{j}_{xx},\cdots,\bar{w}^{j}_{yy},\cdots\}$, where $(i=1,2,\cdots,N\times N; \,\, j=1,2,\cdots,4N)$. A  quadrature element for a gradient Kirchhoff plate with $N_{x} \times N_{y}$ grid is shown in the Figure \ref{fig:Plate}.

Here, $N=N_{x}=N_{y}=5$ are the number of grid points in $\xi$ and $\eta$ directions, respectively. It can be seen from the Figure \ref{fig:Plate}, the plate element has three degrees of freedom on each edge, five degrees of freedom at the corners and only displacement in the domain. The slope $\bar{w}^{'}$ and curvature $\bar{w}^{''}$ dofs related to each edge of the plate are highlighted by the boxes. The transformation used for the plate is $\xi=2x/l_{x}$ and $\eta=2y/l_{y}$ with $-1\leq (\xi,\eta)\leq 1$. 

\subsection{Lagrange interpolation based quadrature element for gradient elastic plates}\label{section_Lag_qem_plate}
The displacement for a $N_{x} \times N_{y}$ node quadrature	 plate element is assumed as
\begin{align*}    \label{disp_Lag_qem_plate}
\bar{w}(x,y,t)=\sum_{i=1}^{N} \sum_{j=1}^{N}L_{i}(x)L_{j}(y){w}^{p}_{ij}(t)=\sum_{i=1}^{N} \sum_{j=1}^{N} \bar{L}_{i}(\xi)\bar{L}_{j}(\eta){w}^{p}_{ij}(t)\tag{48}
\end{align*}

\noindent where $\bar{w}^{p}_{ij}(t)$ is the nodal displacement vector for the plate and $\bar{L}_{i}(\xi)$, $\bar{L}_{j}(\eta)$ are the Lagrange interpolation functions in $\xi$ and $\eta$ directions, respectively. The slope and curvature degrees of freedom at the element boundaries are accounted while computing the weighting coefficients of higher order derivatives as discussed in section \ref{section_Lag_qem_beam}. Substituting the above Equation (\ref{disp_Lag_qem_plate}) in Equation (\ref{K_plate}) we get the stiffness matrix for a gradient elastic quadrature plate element as
 \begin{align*}      \label{Lag_Kij_dqm_plate_cl}         
[K]_{cl}=\frac{ab}{4}\sum_{i=1}^{N}\sum_{j=1}^{N} {H}_{i}{H}_{j}\big[{F}(\xi_{i},\eta_{j})\big]_{cl}^{T}\,[D]_{cl}\,[{F}(\xi_{i},\eta_{j})]_{cl} \tag{49}
\end{align*}
\begin{align*}      \label{Lag_Kij_dqm_plate_ncl}         
[K]_{sg}=g^{2}\frac{ab}{4}\sum_{i=1}^{N}\sum_{j=1}^{N} {H}_{i}{H}_{j}[{F}(\xi_{i},\eta_{j})]_{sg}^{T}\,[D]_{sg}\,[{F}(\xi_{i},\eta_{j})]_{sg}\tag{50}
\end{align*}

\noindent where ($\xi_{i}$ $\eta_{j})$ and $(H_{i}, H_{j})$ are abscissas and weights of GLL quadrature rule. $[F(\xi_{i},\eta_{j})]_{cl}$ and $[F(\xi_{i},\eta_{j})]_{sg}$ are the classical and non-classical strain matrices  at location $(\xi_{i} , \eta_{j})$ for gradient elastic plate. $[D]_{cl}$ and $[D]_{sg}$ are the constitutive matrices corresponding to classical and gradient elastic plate. The classical and non-classical strain matrices are defined as
\begin{align*} \label{Lag_strain_mat_cl_dqm_plate}
\big[{F}(\xi_{i},\eta_{j})\big]_{cl}\{\bar{w}^{p}\}=\begin{bmatrix}
\mathlarger{\frac{4}{a^{2}}}\mathlarger{\mathlarger{‎‎\sum}}_{k=1}^{N+4}\bar{B}_{ik}^{\xi}\bar{w}^{p}_{kj} \\ \\ 
\mathlarger{\frac{4}{b^{2}}}\mathlarger{\mathlarger{‎‎\sum}}_{k=1}^{N+4}\bar{B}_{ik}^{\eta}\bar{w}^{p}_{ik} \\ \\ 
\mathlarger{\frac{8}{ab}}\mathlarger{\mathlarger{‎‎\sum}}_{l=1}^{N+4}\mathlarger{\mathlarger{‎‎\sum}}_{k=1}^{N+4}\bar{A}_{il}^{\xi}\bar{A}_{jk}^{\eta}\bar{w}^{p}_{lk}  
\end{bmatrix} \,\,\,\,\,\, (i,j=1,2,..,N)\tag{51}
\end{align*}
\begin{align*} \label{Lag_strain_mat_ncl_dqm_plate}
\big[{F}(\xi_{i},\eta_{j})\big]_{sg}\{\bar{w}^{p}\}=\begin{bmatrix} 
\mathlarger{g^{2}\frac{8}{a^{3}}}\mathlarger{\mathlarger{‎‎\sum}}_{k=1}^{N+4}\bar{C}_{ik}^{\xi}\bar{w}^{p}_{kj} \\ \\ 
\mathlarger{g^{2}\frac{8}{b^{3}}}\mathlarger{\mathlarger{‎‎\sum}}_{k=1}^{N+4}\bar{C}_{ik}^{\eta}\bar{w}^{p}_{ik} \\ \\ 
\mathlarger{g^{2}\frac{8}{{a}^{2}b}}\mathlarger{\mathlarger{‎‎\sum}}_{l=1}^{N+4}\mathlarger{\mathlarger{‎‎\sum}}_{k=1}^{N+4}\bar{B}_{il}^{\xi}\bar{A}_{jk}^{\eta}\bar{w}^{p}_{lk}  
\\ \\ 
\mathlarger{g^{2}\frac{8}{a{b}^{2}}}\mathlarger{\mathlarger{‎‎\sum}}_{l=1}^{N+4}\mathlarger{\mathlarger{‎‎\sum}}_{k=1}^{N+4}\bar{A}_{il}^{\xi}\bar{B}_{jk}^{\eta}\bar{w}^{p}_{lk}  
 \end{bmatrix} \,\,\,\,\,\, (i,j=1,2,..,N)\tag{52}
\end{align*} 

The classical and non-classical constitutive matrices are given as
\begin{align*} \label{Lag_constitutive_mat_cl_dqm_plate}
\big[D]_{cl}=\begin{bmatrix}
\,1 & \phantom{-}\mu & \phantom{-}0 \\ \\ 
\,\mu & \phantom{-}1 & \phantom{-}0 \\ \\ 
\,0 & \phantom{-}0 & \phantom{-}2(1-\mu) \end{bmatrix} \tag{53}
\end{align*}
\begin{align*}  \label{Lag_constitutive_mat_ncl_dqm_plate}
\big[D]_{sg}=\begin{bmatrix}
\,1 & \phantom{-}0 & \phantom{-}\mu & \phantom{-}0 \\ \\ 
\,0  & \phantom{-}1 & \phantom{-}0  & \phantom{-}\mu \\ \\ 
\,0  & \phantom{-}\mu & \phantom{-}(3-2\mu) & \phantom{-}0 \\ \\
\,0  & \phantom{-}\mu & \phantom{-}0 & \phantom{-}(3-2\mu) \end{bmatrix} \tag{54}
\end{align*}

The diagonal mass matrix is given by
\begin{align*}     \label{Lag_Mij_dqm_plate}         
M_{kk}=\frac{\rho{h}ab}{4}H_{i}H{j}\,\,\,\,\,\,\,\, (i,j=1,2,...,N)\,\,\, (k=(i-1)\times N +j)\tag{55}
\end{align*}

\subsection{Mixed interpolation based quadrature element for gradient elastic plates} \label{section_Herm_qem_plate}
The quadrature element presented here is based on mixed Lagrange-Hermite interpolation, with Lagrangian interpolation is assumed in $\xi$ direction and Hermite interpolation in $\eta$ direction. The displacement for a $N_{x} \times N_{y}$ node mixed interpolation quadrature plate element is assumed as
\begin{align*}    \label{disp_Herm_qem_plate}
\bar{w}(x,y,t)=\sum_{i=1}^{N} \sum_{j=1}^{N+4}L_{i}(x)\Gamma_{j}(y){w}^{p}_{ij}(t)=\sum_{i=1}^{N} \sum_{j=1}^{N+4} \bar{L}_{i}(\xi)\bar{\Gamma}_{j}(\eta){w}^{p}_{ij}(t)\tag{56}
\end{align*}

\noindent where $\bar{w}^{p}_{ij}(t)$ is the nodal displacement vector of the plate and $\bar{L}_{i}(\xi)$ and $\bar{\Gamma}_{j}(\eta)$ are the Lagrange and Hermite interpolation functions in $\xi$ and $\eta$ directions, respectively. The formulations based on mixed interpolation methods have advantage in excluding the mixed derivative dofs at the free corners of the plate\cite{Wangb}. The modified weighting coefficient matrices derived in section \ref{section_Lag_qem_beam}, using Lagrange interpolations and those given in section \ref{section_Herm_qem_beam}, for Hermite interpolations are used in forming the element matrices. Substituting the above Equation (\ref{disp_Herm_qem_plate}) in Equation (\ref{K_plate}), we get the stiffness matrix for gradient elastic quadrature plate element based on mixed interpolation as
 \begin{align*}         \label{Herm_Kij_dqm_plate}             
[K]_{cl}=\frac{ab}{4}\sum_{i=1}^{N}\sum_{j=1}^{N} {H}_{i}{H}_{j}\big[{G}(\xi_{i},\eta_{j})\big]_{cl}^{T}\,[D]_{cl}\,[{G}(\xi_{i},\eta_{j})]_{cl} \tag{57}
\end{align*}
\begin{align*}         \label{Herm_Kij_dqm_plate}             
[K]_{sg}=g^{2}\frac{ab}{4}\sum_{i=1}^{N}\sum_{j=1}^{N} {H}_{i}{H}_{j}[{G}(\xi_{i},\eta_{j})]_{sg}^{T}\,[D]_{sg}\,[{G}(\xi_{i},\eta_{j})]_{sg} \tag{58}
\end{align*}

\noindent where ($\xi_{i}$ $\eta_{j})$ and $(H_{i}, H_{j})$ are abscissas and weights of GLL quadrature rule. $[D]_{cl}$ and $[D]_{sg}$ are the classical and gradient elastic constitutive matrices for the plate defined in the section \ref{section_Lag_qem_plate}. The classical  $[G(\xi_{i},\eta_{j})]_{cl}$  and non-classical $[G(\xi_{i},\eta_{j})]_{sg}$ strain matrices at the location $(\xi_{i} , \eta_{j})$ are defined as,.
 \begin{align*} \label{Herm_strain_mat_cl_dqm_plate}
\big[{G}(\xi_{i},\eta_{j})\big]_{cl}\{\bar{w}^{p}\}=\begin{bmatrix}
\mathlarger{\frac{4}{a^{2}}}\mathlarger{\mathlarger{‎‎\sum}}_{k=1}^{N+4}\bar{B}_{ik}^{(\xi)}\bar{w}^{p}_{kj} \\ \\ 
\mathlarger{\frac{4}{b^{2}}}\mathlarger{\mathlarger{‎‎\sum}}_{k=1}^{N+4}\bar{\Gamma}_{jk}^{2(\eta)}\bar{w}^{p}_{ik} \\ \\ 
\mathlarger{\frac{8}{ab}}\mathlarger{\mathlarger{‎‎\sum}}_{l=1}^{N+4}\mathlarger{\mathlarger{‎‎\sum}}_{k=1}^{N+4}\bar{A}_{il}^{(\xi)}\bar{\Gamma}_{jk}^{1(\eta)}\bar{w}^{p}_{lk}  
\end{bmatrix} \,\,\,\,\,\, (i,j=1,2,..,N)\tag{59}
\end{align*}
\begin{align*} \label{Herm_strain_mat_ncl_dqm_plate}
\big[{F}(\xi_{i},\eta_{j})\big]_{sg}\{\bar{w}^{p}\}=\begin{bmatrix} 
\mathlarger{g^{2}\frac{8}{a^{3}}}\mathlarger{\mathlarger{‎‎\sum}}_{k=1}^{N+4}\bar{C}_{ik}^{(\xi)}\bar{w}^{p}_{kj} \\ \\ 
\mathlarger{g^{2}\frac{8}{b^{3}}}\mathlarger{\mathlarger{‎‎\sum}}_{k=1}^{N+4}\bar{\Gamma}_{jk}^{3(\eta)}\bar{w}^{p}_{ik} \\ \\ 
\mathlarger{g^{2}\frac{8}{{a}^{2}b}}\mathlarger{\mathlarger{‎‎\sum}}_{l=1}^{N+4}\mathlarger{\mathlarger{‎‎\sum}}_{k=1}^{N+4}\bar{\Gamma}_{il}^{2(\xi)}\bar{A}_{jk}^{(\eta)}\bar{w}^{p}_{lk}  
\\ \\ 
\mathlarger{g^{2}\frac{8}{a{b}^{2}}}\mathlarger{\mathlarger{‎‎\sum}}_{l=1}^{N+4}
\mathlarger{\mathlarger{‎‎\sum}}_{k=1}^{N+4}
\bar{\Gamma}_{il}^{1(\xi)}\bar{B}_{jk}^{(\eta)}\bar{w}^{p}_{lk}  
 \end{bmatrix} \,\,\,\,\,\, (i,j=1,2,..,N)\tag{60}
\end{align*}

The diagonal mass matrix remains the same as Equation (\ref{Lag_Mij_dqm_plate}). Here, $\bar{A}$, $\bar{B}$ and $\bar{C}$ are the first, second and third order derivatives of Lagrange interpolation functions along the $\xi$ direction. Similarly, $\bar{\Gamma}^{1}$, $\bar{\Gamma}^{2}$ and $\bar{\Gamma}^{3}$ are the first, second and third order derivatives of Hermite interpolation functions in the $\eta$ direction .

\section{Numerical Results and Discussion}
 
The efficiency of the proposed quadrature beam and plate element is demonstrate through free vibration analysis. Initially, the convergence study is performed for an Euler-Bernoulli gradient beam, followed by frequency comparisons for different boundary conditions and $g$ values. Similar, study is conducted for a Kirchhoff plate and the numerical results are tabulated and compared with available literature. Four different values of length scale parameters, $g=0.00001, 0.05, 0.1$, and $0.5$ are considered in this study. Single element is used with GLL quadrature points as nodes to generate all the results reported herein. For results comparison the proposed gradient quadrature beam element based on Lagrange interpolation is designated as SgQE-L and the element based on Hermite interpolation as SgQE-H. Similarly, the plate element based on Lagrange interpolation in $\xi$ and $\eta$ directions as SgQE-LL and the element based on mixed interpolation as SgQE-LH. In this study, the rotary inertia related to slope and curvature degrees of freedom is neglected.

\subsection{Quadrature beam element for gradient elasticity theory}

The numerical data used for the analysis of beams is as follows: Length $L=1$, Young's modulus $E=3 \times 10^{6}$, Poission's ratio $\nu=0.3$ and density $\rho=1$. All the frequencies reported for beams are nondimensional as $\bar\omega=\omega{L^{2}\sqrt{\rho{A}/EI}}$. Where $A$ and $I$ are area and moment of inertia  of the beam and $\omega$ is the natural frequency. The analytical solutions for gradient elastic Euler-Bernoulli beam with different boundary conditions are obtained by following the approach given in \cite{Kitar} and the associated frequency equations are presented in Appendix-I. The  classical and non-classical boundary conditions used in the free vibration analysis for different end support are:\\

\noindent \text{Simply supported} :\\
\noindent \textit{classical} :\,\,$w=M=0$ , \,\,\,\textit{non-classical} : $w^{''}=0$ \hspace{0.2cm} at $x=(-\frac{L}{2},\frac{L}{2})$ \\ 

\noindent \text{Clamped} :\\
\noindent \textit{classical} :\,\,$w=w^{'}=0$ , \,\,\,\textit{non-classical} : $w^{''}=0$ \hspace{0.2cm} at $x=(-\frac{L}{2},\frac{L}{2})$ \\

\noindent \text{Free-free} :\\
\noindent \textit{classical} :\,\,$Q=M=0$ , \,\,\,\textit{non-classical} : $\bar{M}=0$ \hspace{0.2cm} at $x=(-\frac{L}{2},\frac{L}{2})$ \\

\noindent \text{Cantilever} :\\
\noindent \textit{classical} : $w=w^{'}=0$  \hspace{0.2cm} at\hspace{0.1cm}  $x=-\frac{L}{2}$,\hspace{0.2cm} $Q=M=0$ \hspace{0.05cm} at \hspace{0.2cm} $x=\frac{L}{2}$ \,\,\,\,\\
\noindent \textit{non-classical} : $w^{''}=0$ \hspace{0.2cm} at $x=-\frac{L}{2}$ ,\hspace{0.2cm} $\bar{M}=0$ \hspace{0.2cm} at  \hspace{0.1cm} $x=\frac{L}{2}$ \,\,\,\,\\

\noindent \text{Propped cantilever} :\\
\noindent \textit{classical} : $w=w^{'}=0$\hspace{0.3cm} at $x=-\frac{L}{2}$ ,\hspace{0.2cm} $w=M=0$\hspace{0.3cm} at  \hspace{0.1cm} $x=\frac{L}{2}$ \,\,\,\,\\
\noindent \textit{non-classical} : $w^{''}=0$ \,\, at $x=-\frac{L}{2}$ ,\hspace{0.2cm} $w^{''}=0$ \hspace{0.2cm} at  \hspace{0.1cm} $x=\frac{L}{2}$ \,\,\,\,\\

The size of the displacement vector $\Delta_{d}$ defined in Equation (\ref{eq:Boundary_disp_Beam}) remains as $N-2$ for all the boundary conditions of the beam except for free-free and cantilever beam which are $N$ and $N-1$, respectively. However, the size of the $\Delta_{b}$ vector depends upon the number of non-zero slope and curvature dofs at the element boundaries. The non-classical boundary conditions employed for simply supported gradient beam are $w^{''}=0$ at $x=(-\frac{L}{2},\frac{L}{2})$, the equations related to curvature degrees of freedom are eliminated and the size of $\Delta_{b}$ is 2. For the gradient cantilever beam the non-classical boundary conditions used are $w^{''}=0$ at $x=-\frac{L}{2}$ and $\bar{M}=0$ at $x=\frac{L}{2}$. The equation related to curvature degrees of freedom at $x=-\frac{L}{2}$ is eliminated and the equation related to higher order moment at $x=\frac{L}{2}$ is retained and the size of $\Delta_{b}=2$. In the case of clamped beam the non-classical boundary conditions read $w^{''}=0$ at $x=(-\frac{L}{2},\frac{L}{2})$ and the $\Delta_{b}$ is zero. Similarly, the size for the propped cantilever beam will be 3 as  $w^{''}=0$ at $x=(-\frac{L}{2},\frac{L}{2})$. Finally, for a free-free beam the size of $\Delta_{b}$ vector is 4 due to $\bar{M}=0$ at $x=(-\frac{L}{2},\frac{L}{2})$.

\subsubsection{Frequency convergence for gradient elastic quadrature beam elements} 
 
\begin{figure}[H]
\includegraphics[width=1.0\textwidth]{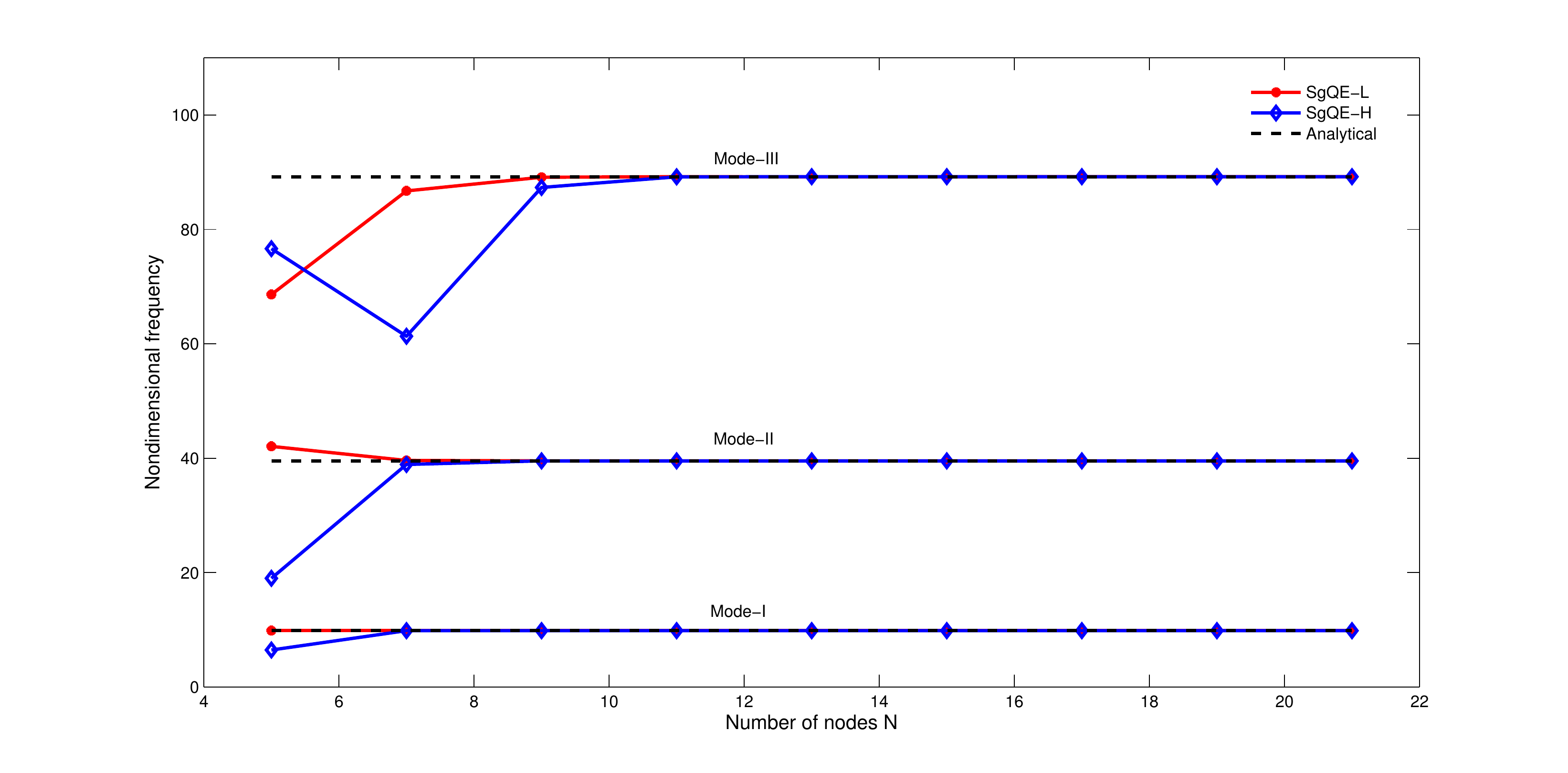}
\centering
\caption{Convergence behaviour of frequencies for a simply supported gradient beam ($g/L=0.1$).}
\label{fig:conv_ss_beam}
\end{figure}

\begin{figure}[H]
\includegraphics[width=1.0\textwidth]{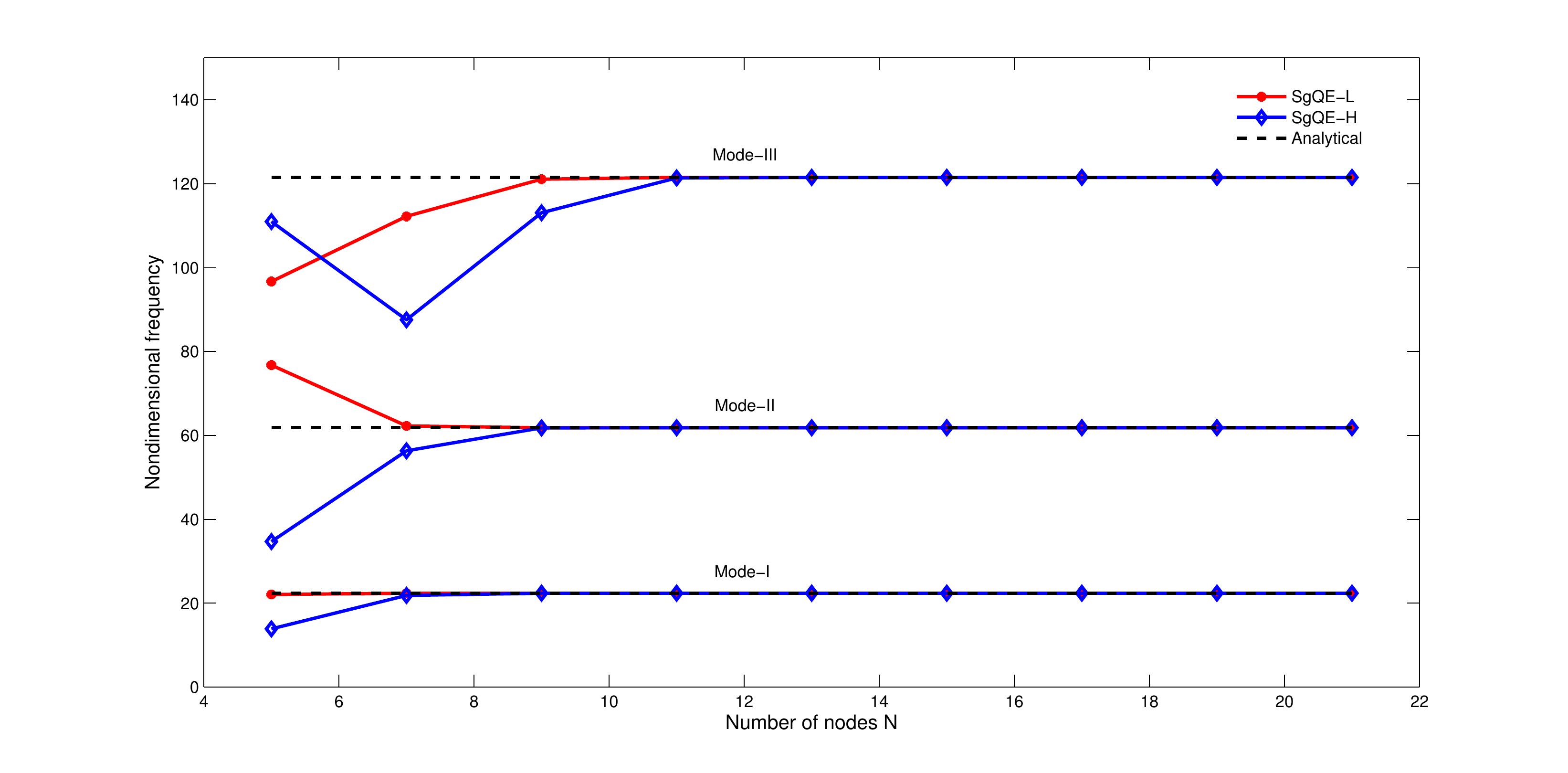}
\centering
\caption{Convergence behaviour of frequencies for a free-free gradient beam ($g/L=0.1$).}
\label{fig:conv_free_beam}
\end{figure}

In this section, the convergence behaviour of frequencies obtained using proposed SgQE-L and SgQE-H elements for simply supported and free-free Euler-Bernoulli beam are compared. Figure \ref{fig:conv_ss_beam}, shows the comparison of first three frequencies for a simply supported gradient beam and their convergence trends for $g/L=0.1$. The convergence is seen faster for both SgQE-L and SgQE-H elements for all the three frequencies with solution converging to analytical values with 10 nodes. Similar trend is noticed in the the Figure \ref{fig:conv_free_beam}, for free-free beam. It is to be noted that, the proposed SgQE-L and SgQE-H elements are efficient in capturing the rigid body modes associated with the generalized degrees of freedom. The frequencies reported for free-free beam are related to elastic modes and the rigid mode frequencies are not reported here, which are zeros. Hence, single SgQE-L or SgQE-H element with fewer number of nodes can produce accurate solutions even for higher frequencies.

\subsubsection{Free vibration analysis of gradient beams using SgQE-L and SgQE-H elements}
To demonstrate the applicability of the  SgQE-L and SgQE-H elements for different boundary conditions the frequencies are compared with the analytical solutions in Tables \ref{ss_freq}-\ref{prop_freq}. The comparison is made for first six frequencies obtained for different values of $g/L=0.00001, 0.05, 0.1, 0.5$. 

\setlength{\extrarowheight}{.5em}
\begin{table}[H]
    \centering
    \caption{Comparison of first six frequencies for a simply supported gradient beam}
      
 \begin{tabular}{p{2.1em}p{5.1em}c@{\hskip 0.3in}c@{\hskip 0.3in}c@{\hskip 0.3in}c@{\hskip 0.3in}}
     \\ \hline
   $ \text{Freq.}$ & $g/L$      & 0.00001	         & 0.05           & 0.1     & 0.5  \\ \hline 
    	    & SgQE-L        & 9.869    & 9.870  & 9.874 & 9.984            \\
	$\bar{\omega}_{1}$    	 & SgQE-H        & 9.869    & 9.871  & 9.874 & 9.991             \\
                             & Analytical   & 9.870    & 9.871  & 9.874 & 9.991            \\ \hline
                                       & SgQE-L        & 39.478    & 39.498  & 39.554                                       & 41.302           \\
	$\bar{\omega}_{2}$     & SgQE-H     & 39.478    & 39.498  & 39.556                                       & 41.381           \\
                             & Analytical         & 39.478    & 39.498  & 39.556                                       & 41.381            \\ \hline 
                                       & SgQE-L        & 88.826                                          & 88.923  & 89.207                                       & 97.725  \\
	$\bar{\omega}_{3}$     & SgQE-H        & 88.826                                          & 88.925  & 89.220                                       & 98.195             \\
                             & Analytical   & 88.826                                          & 88.925  & 89.220                                       & 98.195             \\  \hline
                                       & SgQE-L        & 157.914    & 158.221  & 159.125                                       & 185.378             \\
	$\bar{\omega}_{4}$                             & SgQE-H         & 157.915    & 158.225  & 159.156                                       & 186.497            \\
                             & Analytical    & 157.914    & 158.226  & 159.156                                       & 186.497           \\  \hline
                                      & SgQE-L       & 246.740    & 247.480  & 249.655                                       & 310.491         \\
	$\bar{\omega}_{5}$                              & SgQE-H          & 247.305    & 247.475  & 249.760                                       & 313.741             \\
                             & Analytical   & 246.740    & 247.500  & 249.765                                       & 313.743            \\
  \hline
                                       & SgQE-L        & 355.344    & 357.039  & 361.805                                       & 486.229             \\
	$\bar{\omega}_{6}$                             & SgQE-H               & 355.306    & 356.766  & 361.564                                       & 488.302     \\
                                                              & Analytical        & 355.306    & 356.880  & 361.563                                       & 488.240           \\  \hline

        \end{tabular}
    \label{ss_freq}
\end{table}

\setlength{\extrarowheight}{.5em}
\begin{table}[H]
    \centering
    \caption{Comparison of first six frequencies for a free-free gradient beam}
      
 \begin{tabular}{p{2.1em}p{5.1em}c@{\hskip 0.3in}c@{\hskip 0.3in}c@{\hskip 0.3in}c@{\hskip 0.3in}}
     \\ \hline
   $ \text{Freq.}$ & $g/L$      & 0.00001	         & 0.05           & 0.1     & 0.5  \\ \hline 
         & SgQE-L        & 22.373    & 22.376  & 22.387                                       & 22.691     \\
	$\bar{\omega}_{1}$  			 & SgQE-H       & 22.373    & 22.377  & 22.387                                       & 22.692          \\
                             & Analytical     & 22.373    & 22.377  & 22.387                                       & 22.692          \\ \hline
	$\bar{\omega}_{2}$        & SgQE-L        & 61.673    & 61.708  & 61.814                                       & 64.841             \\
                               & SgQE-H        & 61.673    & 61.708  & 61.814                                       & 64.856           \\
                             & Analytical      & 61.673    & 61.708  & 61.814                                       & 64.856           \\ \hline 
   & SgQE-L        & 120.903                                          & 121.052  & 121.496                                       & 133.627             \\
	$\bar{\omega}_{3}$          & SgQE-H        & 120.904    & 121.052  & 121.497                                  & 133.710   \\
                             & Analytical      & 120.903    & 121.052  & 121.497                                  & 133.710          \\  \hline
        & SgQE-L        & 199.859    & 202.864  & 201.553                                       & 234.596           \\
	$\bar{\omega}_{4}$       & SgQE-H        & 199.876    & 200.287  & 201.556                                       & 234.875          \\   
                             & Analytical     & 199.859    & 200.286  & 201.557                                       & 234.875           \\     \hline
          & SgQE-L        & 298.550    & 299.528  & 302.422                                       & 374.535     \\
	$\bar{\omega}_{5}$          & SgQE-H        & 298.556    & 299.365  & 302.403                                       & 375.234     \\   
                             &Analytical        & 298.555    & 299.537  & 302.443                                       & 375.250     \\   
  \hline
     & SgQE-L        & 417.217    & 419.418  & 425.469                                       & 562.869     \\
	$\bar{\omega}_{6}$           & SgQE-H        & 416.991    & 418.438  & 424.747                                       & 562.758    \\   
  &Analytical       & 416.991    & 418.942  & 424.697                                       & 562.536     \\   
  \hline

        \end{tabular}
    \label{free_freq}
\end{table}

 In the Table \ref{ss_freq}, the comparison of first six frequencies for a simply supported gradient beam are shown. For $g/L=0.00001$, all the frequencies match well with the exact frequencies of classical beam. Good agreement with analytical solutions is noticed for all the frequencies obtained using SgQE-L and SgQE-H elements for higher values of $g/L$. In Table \ref{free_freq}, the frequencies corresponding to elastic modes are tabulated and compared for a free-free beam. Similarly, in Tables \ref{clamped_freq}-\ref{prop_freq}, comparison in made for cantilever, clamped and propped cantilever gradient beams, respectively. The frequencies obtained using SgQE-L and SgQE-H elements are in close agreement with the analytical solutions for different $g/L$ values. Hence, based on the above findings it can be stated that the SgQE-I and SgQE-II elements can be applied for free vibration analysis of gradient Euler-Bernoulli beam for any choice of boundary conditions and $g/L$ values.

\setlength{\extrarowheight}{.5em}
\begin{table}[H]
    \centering
    \caption{Comparison of first six frequencies for a clamped gradient beam}
      
 \begin{tabular}{p{2.1em}p{5.1em}c@{\hskip 0.3in}c@{\hskip 0.3in}c@{\hskip 0.3in}c@{\hskip 0.3in}}
     \\ \hline
   $ \text{Freq.}$ & $g/L$      & 0.00001	         & 0.05           & 0.1     & 0.5  \\ \hline 
   & SgQE-L        & 22.324    & 22.801  & 23.141                                       & 27.747     \\
	$\bar{\omega}_{1}$      & SgQE-H       & 22.590    & 22.845  & 23.310                                       & 27.976      \\
                             & Analytical      & 22.373    & 22.831  & 23.310                                       & 27.976     \\   \hline
                             & SgQE-L        & 61.540    & 62.720  & 63.984                                       & 79.450      \\
	$\bar{\omega}_{2}$          & SgQE-H        & 62.276    & 63.003  & 64.365                                       & 79.970     \\
                             & Analytical   & 661.673    & 62.961  & 64.365                                       & 79.970       \\  \hline
              & SgQE-L        & 120.392    & 122.916	  & 	125.542                                       & 162.889       \\
	$\bar{\omega}_{3}$       & SgQE-H        & 122.094    & 123.594	  & 	126.512                                       & 164.927          \\
                             & Analytical  & 120.903    & 123.511	  & 	126.512                                       & 164.927   \\  \hline
                 & SgQE-L       & 199.427    & 203.581  & 208.627 & 286.576            \\   
	$\bar{\omega}_{4}$       & SgQE-H        & 201.843    & 204.502  & 209.887 & 289.661     \\   
                             & Analytical   & 199.859    & 204.356  & 209.887 & 289.661   \\     \hline
             & SgQE-L       & 297.282    & 304.138  & 312.503                                       & 455.285      \\   
	$\bar{\omega}_{5}$   & SgQE-H        & 301.541    & 305.843  & 314.956                                       & 462.238     \\   
                             & Analytical   & 298.555    & 305.625  & 314.956                                       & 462.238      \\     \hline
    & SgQE-L        & 421.194    & 427.786  & 442.299                                       & 681.749       \\   
	$\bar{\omega}_{6}$        & SgQE-H        & 421.092    & 427.787  & 442.230                                       & 691.292         \\   
  & Analytical  & 416.991    & 427.461  & 442.230                                       & 691.292         \\
    \hline
        \end{tabular}
    \label{clamped_freq}
\end{table}

\setlength{\extrarowheight}{.5em}
\begin{table}[H]
    \centering
    \caption{Comparison of first six frequencies for a cantilever gradient beam}
      
 \begin{tabular}{p{2.1em}p{5.1em}c@{\hskip 0.3in}c@{\hskip 0.3in}c@{\hskip 0.3in}c@{\hskip 0.3in}}
     \\ \hline
   $ \text{Freq.}$ & $g/L$      & 0.00001	         & 0.05           & 0.1     & 0.5  \\ \hline 
  & SgQE-L      & 3.532    & 3.545  & 3.584 & 3.857         \\
	$\bar{\omega}_{1}$      & SgQE-H       & 3.534    & 3.552  & 3.587                                       & 3.890         \\
                             & Analytical      & 3.532    & 3.552  & 3.587                                       & 3.890         \\ \hline
            & SgQE-L       & 21.957    & 22.188  & 22.404                                       & 24.592          \\
	$\bar{\omega}_{2}$ & SgQE-H        & 22.141    & 22.267  & 22.497                                       & 24.782          \\
                             & Analytical  & 22.141    & 22.267  & 22.496                                       & 24.782            \\ \hline 
                & SgQE-L       & 61.473    & 62.150  & 62.822	                                      & 71.207            \\
	$\bar{\omega}_{3}$   & SgQE-H        & 61.997    & 62.375  & 63.094                                      & 71.863         \\
                             & Analytical  & 61.997    & 62.375  & 63.094                                      & 71.863        \\  \hline
        & SgQE-L       & 120.465    & 121.867  & 123.424                                       & 146.652         \\   
	$\bar{\omega}_{4}$   & SgQE-H        & 121.495    & 122.313  & 123.966                                       & 148.181        \\   
                             & Analytical   & 121.495    & 122.313  & 123.966                                       & 148.181         \\     \hline
       & SgQE-L        & 199.141    & 201.636  & 204.752                                       & 257.272         \\   
	$\bar{\omega}_{5}$                & SgQE-H       & 200.848  & 202.377    & 205.658  & 260.336     \\   
                             & Analytical   & 202.377    & 205.658  & 260.336                                       & 200.847      \\   
 \hline
& SgQE-L        & 297.489    & 301.551  & 307.229                                       & 410.222        \\   
	$\bar{\omega}_{6}$      & SgQE-H    & 300.043     & 302.667    & 308.605  & 415.802      \\   
  & Analytical        & 300.043     & 302.667    & 308.605  & 415.802      \\  
  \hline

        \end{tabular}
    \label{cant_freq}
\end{table}

 \setlength{\extrarowheight}{.5em}
\begin{table}[H]
    \centering
    \caption{Comparison of first six frequencies for a propped cantilever gradient beam}
      
 \begin{tabular}{p{2.1em}p{5.1em}c@{\hskip 0.3in}c@{\hskip 0.3in}c@{\hskip 0.3in}c@{\hskip 0.3in}}
     \\ \hline
   $ \text{Freq.}$ & $g/L$      & 0.00001	         & 0.05           & 0.1     & 0.5  \\ \hline 
    & SgQE-L        & 15.413    & 15.520  & 15.720                                       & 17.351 \\
	$\bar{\omega}_{1}$    	 & SgQE-H       & 15.492    & 15.581	  & 15.740                                       & 17.324	        \\
                             & Analytical      & 15.492    & 15.581	  & 15.740                                       & 17.324	        \\ \hline
           & SgQE-L     & 49.869    & 50.313  & 51.026                                       & 57.767         \\
	$\bar{\omega}_{2}$           & SgQE-H        & 50.207    & 50.512  & 51.089                                       & 58.197        \\
       & Analytical    & 50.207    & 50.512  & 51.089                                       & 58.197          \\ \hline 
              & SgQE-L       & 104.044    & 105.043  & 106.557                                       & 127.127         \\
	$\bar{\omega}_{3}$  & SgQE-H        & 104.758    & 105.457  & 106.865                                       & 128.005         \\
                             & Analytical   & 104.758    & 105.457  & 106.865                                       & 128.005          \\ \hline
             & SgQE-L     & 177.922    & 179.778  & 182.822                                       & 231.247         \\   
	$\bar{\omega}_{4}$      & SgQE-H        & 179.149    & 180.500  & 183.389                                       & 233.357         \\   
                             & Analytical      & 179.149    & 180.500  & 183.389                                       & 233.357          \\     \hline
            & SgQE-L      & 271.502    & 274.654  & 280.210                                       & 378.692          \\
	$\bar{\omega}_{5}$       & SgQE-H        & 273.383    & 275.749  & 281.089                                       & 382.058         \\   
                             & Analytical    & 273.383    & 275.749  & 281.089                                       & 382.058          \\   
  \hline
              & SgQE-L        & 384.785    & 389.746  & 399.154                                       & 575.841        \\   
	$\bar{\omega}_{6}$           & SgQE-H        & 387.463    & 391.341  & 400.509                                       & 582.607        \\   
  & Analytical       & 387.463    & 391.341  & 400.509                                       & 582.607        \\     \hline

        \end{tabular}
    \label{prop_freq}
\end{table}

\subsection{Quadrature plate element for gradient elasticity theory}

Three different boundary conditions of the plate, simply supported on all edges (SSSS), free on all edges (FFFF) and combination of simply supported and free (SSFF) are considered. The converge behaviour of SgQE-LL and SgQE-LH plate elements is verified first, later numerical comparisons are made for all the three plate conditions for various $g/l_{x}$ values. All the frequencies reported herein for plate are non-dimensional as $\bar\omega=\omega{l^{2}_{x}\sqrt{\rho{h}/D}}$. The numerical data used for the analysis of plates is: length $l_{x}=1$, width $l_{y}=1$, thickness $h=0.01$, Young's modulus $E=3 \times 10^{6}$, Poission's ratio $\nu=0.3$ and density $\rho=1$. The number of nodes in either direction are assumed to be equal, $N=N_x=N_y$. The choice of the essential and natural boundary conditions for the above three plate problems are given in section \ref{section_Sg_Plate}.

The size of the displacement vector $\Delta_{d}$ defined in Equation (\ref{eq:Boundary_disp_Beam}) remains as $(N-2)\times (N-2)$ for all the boundary conditions of the gradient plate except for free-free and cantilever plate which are $N\times N$ and $(N\times N)-N$, respectively. However, the size of the $\Delta_{b}$ vector depends upon the number of non-zero slope and curvature dofs along the element boundaries. The non-classical boundary conditions employed for SSSS gradient plate are $\bar{w}_{xx}=0$ at $x=(-\frac{l_{x}}{2},\frac{l_{x}}{2})$  and  $\bar{w}_{yy}=0$ at $y=(-\frac{l_{y}}{2},\frac{l_{y}}{2})$, the equations related to curvature degrees of freedom are eliminated and the size of $\Delta_{b}$ will be $4N-8$, as the $\bar{w}_{x}=\bar{w}_{y}=0$ at the corners of the plate. For a FFFF plate the non-classical boundary conditions employed are $\bar{M}_{x}=0$ at $x=(-\frac{l_{x}}{2},\frac{l_{x}}{2})$ and  $\bar{M}_{y}=0$ at $y=(-\frac{l_{y}}{2},\frac{l_{y}}{2})$, and the size of $\Delta_{b}$ is $8N$.  Finally, for SSFF plate $\Delta_{b}=6N-4$.

\subsubsection{Frequency convergence of gradient elastic quadrature plate elements} 
 
\begin{figure}[H]
\includegraphics[width=0.95\textwidth]{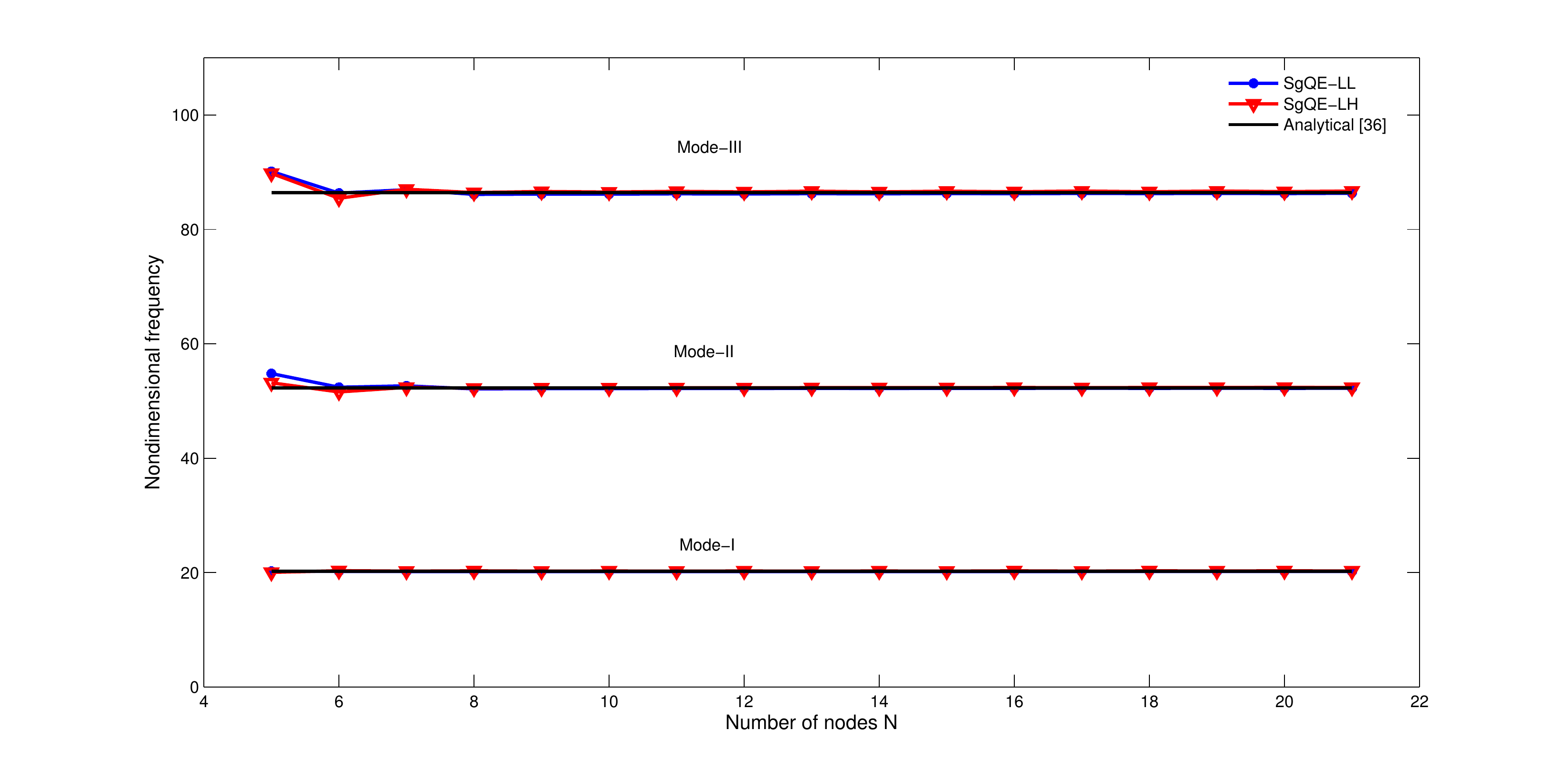}
\centering
\caption{Convergence behaviour of frequencies for a SSSS gradient plate ($g/l_{X}=0.05$).}
\label{fig:conv_SSSS_plate}
\end{figure}

\begin{figure}[H]
\includegraphics[width=0.95\textwidth]{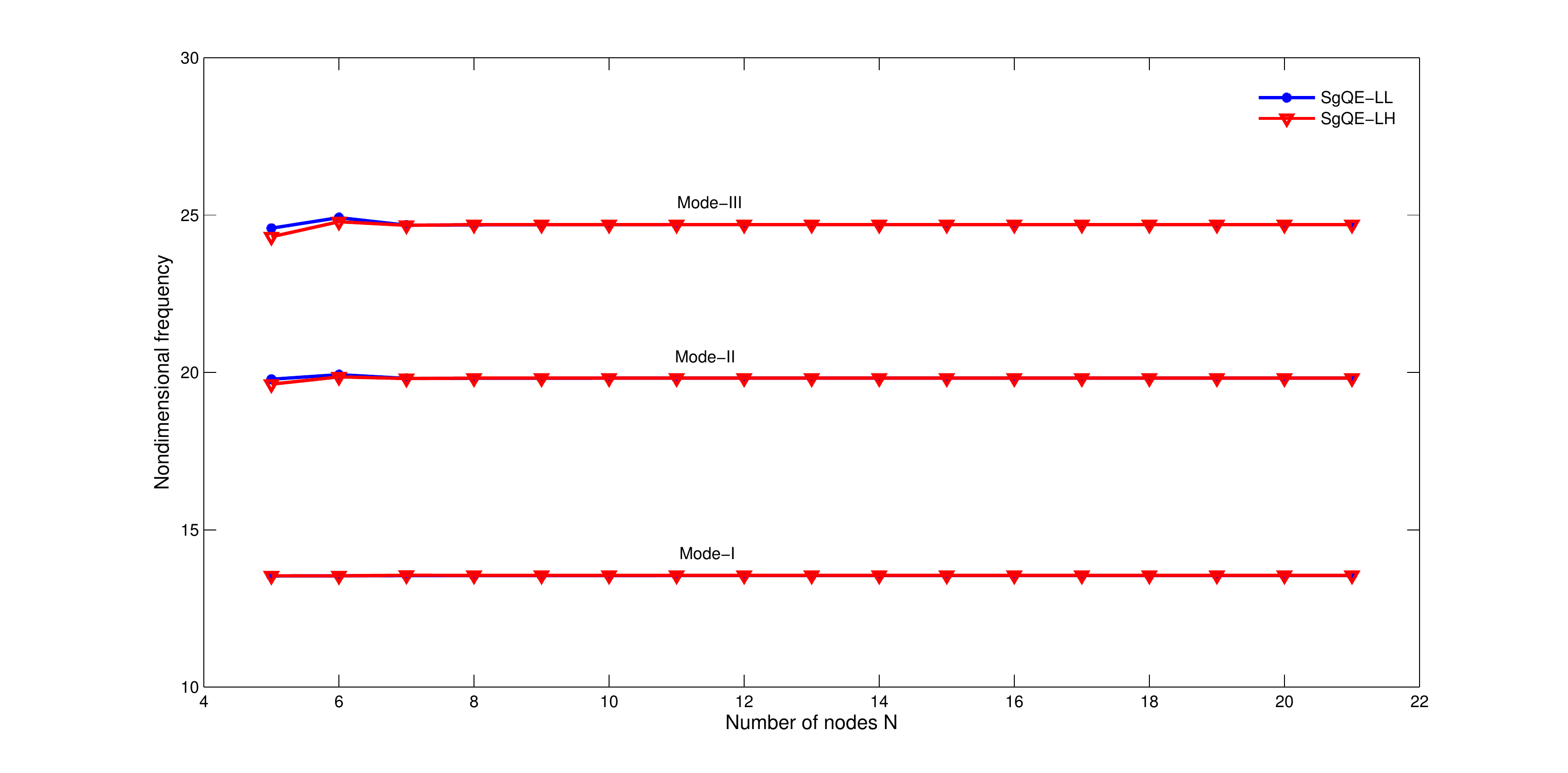}
\centering
\caption{Convergence behaviour of frequencies for a FFFF gradient plate ($g/l_{X}=0.05$).}
\label{fig:conv_FFFF_plate}
\end{figure}

\begin{figure}[h]
\includegraphics[width=0.95\textwidth]{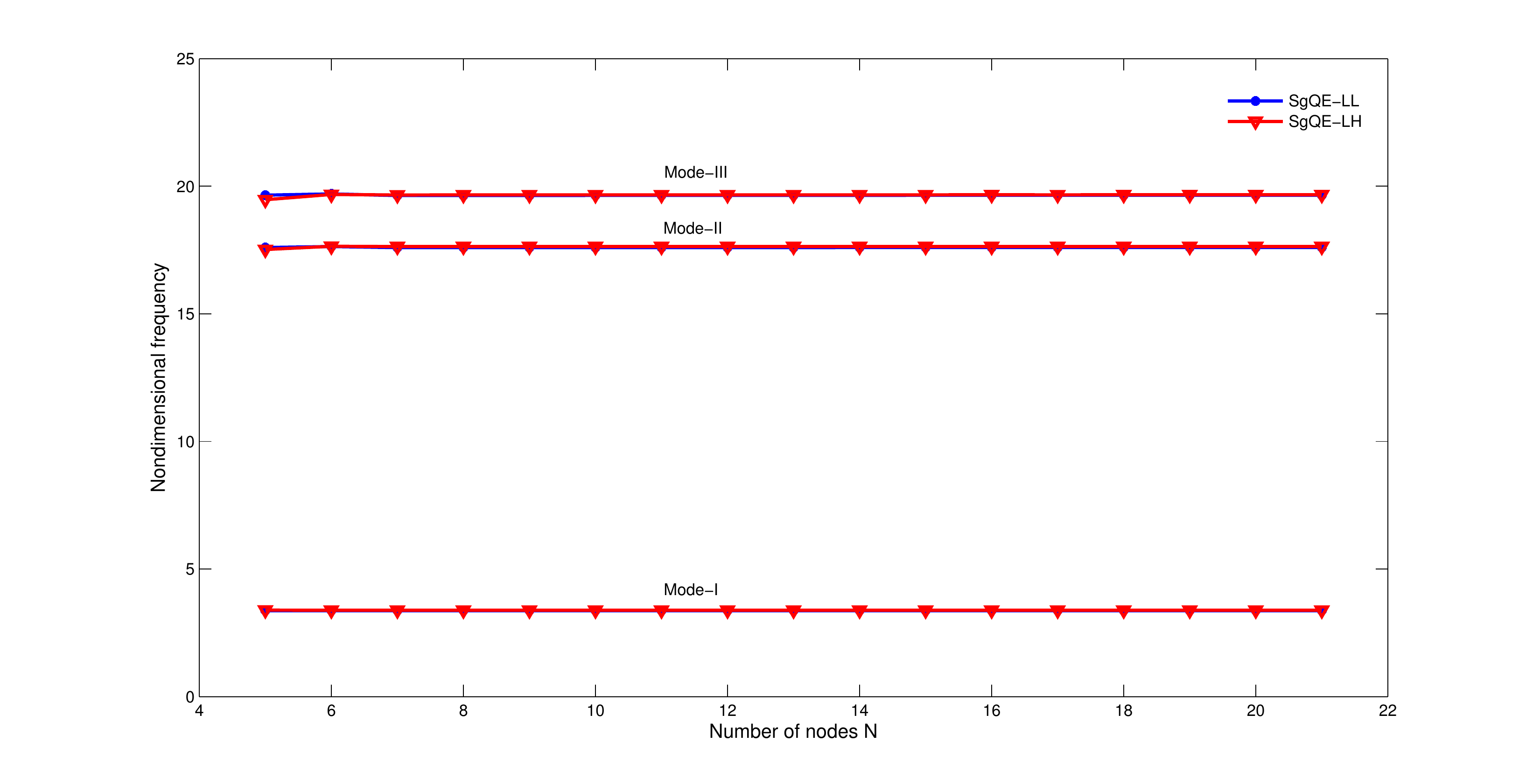}
\centering
\caption{Convergence behaviour of frequencies for a SSFF gradient plate ($g/l_{X}=0.05$).}
\label{fig:conv_SSFF_plate}
\end{figure}

In Figure \ref{fig:conv_SSSS_plate}, convergence of first three frequencies for a SSSS plate obtained using SgQE-LL and SgQE-LH elements for $g/l_{x}=0.05$ is plotted and compared with analytical solutions\cite{Besko1p}. Both SgQE-LL and SgQE-LH elements show excellent convergence behaviour for all the three frequencies. Figures \ref{fig:conv_FFFF_plate} and \ref{fig:conv_SSFF_plate}, illustrate the frequency convergence for FFFF and SSFF plates, respectively, for $g/l_{x}=0.05$. Only the SgQE-LL and SgQE-LH element frequencies are shown, as the gradient solution are not available in literature for comparison. It is observed that SgQE-LL and SgQE-LH elements exhibit identical convergence characteristics.

\subsubsection{Free vibration analysis of gradient plate using SgQE-LL and SgQE-LH elements}
The first six frequencies for SSSS, FFFF and SSFF plates obtained using SgQE-LL and SgQE-LH elements are compared and tabulated. The comparison is made for different length scale parameter: $g/l_{x}=0.00001, 0.05, 0.1, 0.5$. All the tabulated reslts are generated using $N_{x}=N_{y}=11$ nodes. In the Table \ref{Freq_SSSS_Table}, the comparison of first six frequencies for SSSS gradient plate are shown. Good agreement with analytical solutions\cite{Besko1p} is noticed for all the frequencies obtained using SgQE-LL and SgQE-LH elements for different $g/l_{x}$. 

Tables \ref{Freq_FFFF_Table} and \ref{Freq_SSFF_Table} contains the frequency comparison for FFFF and SSFF plates for various $g/l_{x}$ values. As the exact solutions for gradient elastic plate are not available in the literature for FFFF and SSFF support conditions, the frequencies obtained using SgQE-LL and SgQE-LH are compared. Both elements show identical performance for all $g/l_{x}$ values. The frequencies obtained for lower values of $g/l_{x}=0.00001$ match well with the classical plate frequencies for all support conditions. \\
\indent In the above findings, SgQE-LL and SgQE-LH elements demonstrate excellent agreement with analytical results for all frequencies and $g/l_{x}$ values for a SSSS plate. For FFFF and SSFF plates, SgQE-LL and SgQE-LH elements produce similar frequencies for $g/l_{x}$ values considered. Hence, a single SgQE-LL or SgQE-LH element with few nodes can be used efficiently to study the free vibration behaviour of gradient plates with different support conditions and $g/l_{x}$ values.

\setlength{\extrarowheight}{.5em}
\begin{table}[H]
    \centering
    \caption{Comparison of first six frequencies for a gradient SSSS plate}
      
    \begin{tabular}{p{2.1em}p{5.1em}c@{\hskip 0.3in}c@{\hskip 0.3in}c@{\hskip 0.3in}c@{\hskip 0.3in}}
     \\ \hline
   $ \text{Freq.}$ & $g/l_{x}$      & 0.00001	         & 0.05           & 0.1     & 0.5 \\ \hline 
      & SgQE-LL        & 19.739    & 20.212  & 21.567 & 47.693  \\
	$\bar{\omega}_{1}$   		 & SgQE-LH       & 19.739    & 20.286  & 21.791 & 49.940    \\
       & Analyt.\cite{Besko1p}
       (m=1,n=1)  & 19.739    & 20.220  & 21.600 & 48.087  \\  \hline

     & SgQE-LL        & 49.348    & 52.249  & 60.101                                       & 178.418            \\
	$\bar{\omega}_{2}$   & SgQE-LH     & 49.348      & 52.365    & 60.429                                   & 180.895           \\
       & Analyt.\cite{Besko1p}
 (m=1,n=2)        & 49.348    & 52.303  & 60.307                                       & 180.218       \\     \hline

          & SgQE-LL        & 78.957                                          & 86.311  & 105.316              & 357.227       \\
          
	$\bar{\omega}_{3}$    & SgQE-LH        & 78.957                                          & 86.720  & 106.321              & 363.156        \\
       & Analyt.\cite{Besko1p}
    (m=2,n=2)       & 78.957                                          & 86.399  & 105.624      & 359.572       \\   \hline 
        
  & SgQE-LL        & 98.696    & 109.863  & 137.940                                       & 491.447          \\
	$\bar{\omega}_{4}$ & SgQE-LH         & 98.696     & 109.950   & 138.193                                       & 493.131          \\            
       & Analyt.\cite{Besko1p}
 (m=1,n=3)  & 98.696    & 110.201  & 139.121   & 500.088     \\   \hline

  & SgQE-LL        & 128.305    & 147.119  & 192.759                                       & 730.346      \\
           
	$\bar{\omega}_{5}$     & SgQE-LH    & 128.305    & 147.639 & 193.950                                       & 736.599    \\
  
       & Analyt.\cite{Besko1p}
 (m=2,n=3)     & 128.305    & 147.454  & 193.865                                       & 737.906      \\
  \hline 
  
 & SgQE-LL        & 167.783    & 199.133  & 272.173                                       & 1084.136          \\
	$\bar{\omega}_{6}$  & SgQE-LH               & 167.783    & 199.262  & 272.486                                       & 1085.930            \\
       & Analyt.\cite{Besko1p}
(m=1,n=4)        & 167.783    & 199.897  & 274.562                                        & 1099.535       \\    \hline

        \end{tabular}
    \label{Freq_SSSS_Table}
\end{table}

\setlength{\extrarowheight}{.5em}
\begin{table}[H]
    \centering
    \caption{Comparison of first six frequencies for a gradient FFFF plate}
      
 \begin{tabular}{p{2.1em}p{5.6em}c@{\hskip 0.3in}c@{\hskip 0.3in}c@{\hskip 0.3in}c@{\hskip 0.3in}}
     \\ \hline
   $ \text{Freq.}$ & $g/l_{x}$      & 0.00001	         & 0.05           & 0.1     & 0.5  \\ \hline 
         & SgQE-LL        & 13.468    & 13.546  & 13.681 & 14.118  \\
	$\bar{\omega}_{1}$   		 & SgQE-LH       & 13.468    & 13.551  & 13.713 & 15.628  \\
       & Classical\cite{Wangb}
       (${g/l_{x}}$=0)  & 13.468    & ------  & ------ & ------   \\  \hline

     & SgQE-LL        & 19.596    & 19.820  & 20.313                                       & 22.113       \\
	$\bar{\omega}_{2}$   & SgQE-LH     & 19.596    & 19.820  & 20.315                                       & 22.129       \\
       & Classical\cite{Wangb}
       (${g/l_{x}}$=0)  & 19.596    & ------  & ------ & ------   \\  \hline

          & SgQE-LL        & 24.270                                          & 24.699  & 25.681              & 29.745       \\

	$\bar{\omega}_{3}$    & SgQE-LH        & 24.270                                          & 24.700  & 25.686              & 29.785        \\
     & Classical\cite{Wangb}
       (${g/l_{x}}$=0)       & 24.270      & ------  & ------                                                  & ------\\  \hline

          & SgQE-LL        & 34.8001    & 35.780  & 37.929                                       &73.986            \\
            $\omega_{4}$ & SgQE-LH         & 34.8001    & 35.722  & 38.015                                       & 76.161       \\
                              & Classical\cite{Wangb}
       (${g/l_{x}}$=0)        & 34.8001    & ------ & ------                                                   & ------  \\  \hline

           & SgQE-LL        & 61.093    & 64.314  & 71.238                                       & 145.033            \\
	$\bar{\omega}_{5}$     & SgQE-LH          & 61.093    & 64.317  & 71.244                                       & 145.065           \\
  & Classical\cite{Wangb}
       (${g/l_{x}}$=0)     & 61.093    & ------  & ------                                                     & ------\\
  \hline 
  
     & SgQE-LL        & 63.686    & 67.059  & 75.114                                       & 193.940      \\
	$\bar{\omega}_{6}$  & SgQE-LH               & 63.686    & 67.123  & 75.509 & 200.707  \\
 & Classical\cite{Wangb}
       (${g/l_{x}}$=0)        & 63.686    & ------  & ------                                                    & ------ \\  \hline

        \end{tabular}
    \label{Freq_FFFF_Table}
\end{table}

\setlength{\extrarowheight}{.5em}

\begin{table}[H]
    \centering
    \caption{Comparison of first six frequencies for a gradient SSFF plate}
      
 \begin{tabular}{p{2.1em}p{6.9em}c@{\hskip 0.3in}c@{\hskip 0.3in}c@{\hskip 0.3in}c@{\hskip 0.3in}}
     \\ \hline
   $ \text{Freq.}$ & $g/l_{x}$      & 0.00001	         & 0.05           & 0.1     & 0.5  \\ \hline
      & SgQE-LL        & 3.367    & 3.373  & 3.386 & 3.491 \\
	$\bar{\omega}_{1}$   & SgQE-LH       & 3.367    & 3.382  & 3.413 & 3.950  \\
       &Classical\cite{Leissa,Singh}
       (${g/l_{x}}$=0)  & 3.367    & ------  & ------ & ------     \\  \hline

     & SgQE-LL        & 17.316    & 17.598  & 18.370                                       & 32.579         \\
	$\bar{\omega}_{2}$   & SgQE-LH     & 17.316    & 17.634  & 18.474                                       & 33.927            \\
       & Classical\cite{Leissa,Singh}
       (${g/l_{x}}$=0)  & 17.316    & ------  & ------  & ------   \\  \hline

          & SgQE-LL        & 19.292                                          & 19.645  & 20.585              & 35.825   \\         
          
	$\bar{\omega}_{3}$    & SgQE-LH        & 19.292                                          & 19.664	  & 20.649              & 36.852    \\
     & Classical\cite{Leissa,Singh}
       (${g/l_{x}}$=0)  & 19.292     & ------  & ------            & ------\\  \hline

          & SgQE-LL        & 38.211    & 39.671  & 39.162                                       & 105.800           \\
	$\bar{\omega}_{4}$ & SgQE-LH         & 38.211    & 39.775  & 43.851                                       & 109.959      \\
                              & Classical\cite{Leissa,Singh}
       (${g/l_{x}}$=0)        & 38.211    & ------                                       & ------       & ------ \\  \hline

           & SgQE-LL       & 51.035    & 53.714  & 60.400                                       & 153.000           \\
	$\bar{\omega}_{5}$     & SgQE-LH          & 51.035    & 53.739  & 60.493                                       & 153.980         \\
  & Classical\cite{Leissa,Singh}
       (${g/l_{x}}$=0)    & 51.035    & ------                                       & ------            & ------\\
  \hline 
  
     & SgQE-LL        & 53.487    & 56.431  & 63.699                                       & 158.557      \\
	$\bar{\omega}_{6}$  & SgQE-LH               & 53.487    & 56.537  & 64.000                                          & 161.072         \\
 & Classical\cite{Leissa,Singh}
       (${g/l_{x}}$=0)        & 53.487    & ------                                        & ------          & ------ \\  \hline

        \end{tabular}
    \label{Freq_SSFF_Table}
\end{table}

\section{Conclusion}

Two novel versions of weak form quadrature elements for gradient elastic beam theory were proposed. This methodology was extended to construct two new and different quadrature plate elements based on Lagrange-Lagrange and mixed Lagrange-Hermite interpolations. A new way to account for the non-classical boundary conditions associated with the gradient elastic beam and plate theories were introduced and implemented. The capability of the proposed four elements was demonstrated through free vibration analysis. Based on the findings it was concluded that, accurate solutions can be obtained even for higher frequencies and for any choice of length scale parameter using single beam or plate element with fewer number of nodes. The new results reported for gradient plate for different boundary conditions can be a reference for the research in this field.

\section*{APPENDIX}

\subsection{Analytical solutions for free vibration analysis of gradient elastic Euler-Bernoulli beam}

To obtain the natural frequencies of the gradient elastic Euler-Bernoulli beam which is governed by Equation (\ref{EOM_beam}), we assume a solution of the form

\begin{align*}         
w(x,t)=\bar{w}(x){e}^{i\omega{t}}
\end{align*}

\noindent substituting the above solution in the governing equation (\ref{EOM_beam}), we get

\begin{align*}         
\bar{w}^{iv}-g^{2}\bar{w}^{vi}-\frac{\omega^{2}}{\beta^{2}}\bar{w}=0
\end{align*}

\noindent here, $\beta^{2}=EI/m$, and the above equation has the solution of type

\begin{align*}         
\bar{w}(x)=\sum_{j=1}^{6}c_{i}{e}^{k_{i}x}
\end{align*}

\noindent where, $c_{i}$ are the constants of integration which are determined through boundary conditions and the $k_{i}$ are the roots of the characteristic equation 

\begin{align*}         
{k}^{iv}-g^{2}{k}^{vi}-\frac{\omega^{2}}{\beta^{2}}=0
\end{align*}

After applying the boundary conditions listed in section \ref{section_Sg_Beam} we get,

\begin{align*}         
[F(\omega)]\{C\}=\{0\}
\end{align*}

For non-trivial solution, following condition should be satisfied

\begin{align*}         
det[F(\omega)]=0
\end{align*}

The above frequency equation renders all the natural frequencies for a gradient elastic Euler-Bernoulli beam. The following are the frequency equations for different boundary conditions.\\
 
\noindent (a) Simply supported beam :
$$[F(\omega)]=
\begin{bmatrix}
1 & 1 & 1 & 1 & 1 & 1\\
{e}^{(k_{1}L)} & e^{(k_{2}L)} & e^{(k_{3}L)} & e^{(k_{4}L)} & e^{(k_{5}L)} & e^{(k_{6}L)}\\ \\
{k_{1}}^2 & {k_{2}}^2 & {k_{3}}^2 & {k_{4}}^2 & {k_{5}}^2 & {k_{6}}^2\\ \\
k_{1}^{2}{e}^{(k_{1}L)} & k_{2}^{2}{e}^{(k_{2}L)} & k_{3}^{2}{e}^{(k_{3}L)} & k_{4}^{2}{e}^{(k_{4}L)} & k_{5}^{2}{e}^{(k_{5}L)} & k_{6}^{2}{e}^{(k_{6}L)}\\ \\
k_{1}^{4} & k_{2}^{4} & k_{3}^{4} & k_{4}^{4} & k_{5}^{4} & k_{6}^{4}\\ \\
k_{1}^{4}{e}^{(k_{1}L)} & k_{2}^{4}{e}^{(k_{2}L)} & k_{3}^{4}{e}^{(k_{3}L)} & k_{4}^{4}{e}^{(k_{4}L)} & k_{5}^{4}{e}^{(k_{5}L)} & k_{6}^{4}{e}^{(k_{6}L)}\\ \\
\end{bmatrix}
$$

\noindent (b) Cantilever beam :
$$[F(\omega)]=
\begin{bmatrix}
1 & 1 & 1 & 1 & 1 & 1\\
k_{1} & k_{2} & k_{3} & k_{4} & k_{5} & k_{6}\\ \\
{k_{1}}^2 & {k_{2}}^2 & {k_{3}}^2 & {k_{4}}^2 & {k_{5}}^2 & {k_{6}}^2\\ \\
t_{1} & t_{2} & t_{3} & t_{4} & t_{5} & t_{6} \\ \\
p_{1} & p_{2} & p_{3} & p_{4} & p_{5} & p_{6} \\ \\
k_{1}^{3}{e}^{(k_{1}L)} & k_{2}^{3}{e}^{(k_{2}L)} & k_{3}^{3}{e}^{(k_{3}L)} & k_{4}^{3}{e}^{(k_{4}L)} & k_{5}^{3}{e}^{(k_{5}L)} & k_{6}^{3}{e}^{(k_{6}L)}\\ \\
\end{bmatrix}
$$

\noindent (c) clamped beam :
$$[F(\omega)]=
\begin{bmatrix}
1 & 1 & 1 & 1 & 1 & 1\\
k_{1} & k_{2} & k_{3} & k_{4} & k_{5} & k_{6}\\ \\
{k_{1}}^2 & {k_{2}}^2 & {k_{3}}^2 & {k_{4}}^2 & {k_{5}}^2 & {k_{6}}^2\\ \\

{e}^{(k_{1}L)} & {e}^{(k_{2}L)} & {e}^{(k_{3}L)} & {e}^{(k_{4}L)} & {e}^{(k_{5}L)} & {e}^{(k_{6}L)}\\ \\

k_{1}{e}^{(k_{1}L)} & k_{2}{e}^{(k_{2}L)} & k_{3}{e}^{(k_{3}L)} & k_{4}{e}^{(k_{4}L)} & k_{5}{e}^{(k_{5}L)} & k_{6}{e}^{(k_{6}L)}\\ \\

k_{1}^{2}{e}^{(k_{1}L)} & k_{2}^{2}{e}^{(k_{2}L)} & k_{3}^{2}{e}^{(k_{3}L)} & k_{4}^{2}{e}^{(k_{4}L)} & k_{5}^{2}{e}^{(k_{5}L)} & k_{6}^{2}{e}^{(k_{6}L)}\\ \\

\end{bmatrix}
$$

\noindent (d) Propped cantilever beam :
$$[F(\omega)]=
\begin{bmatrix}
1 & 1 & 1 & 1 & 1 & 1\\
k_{1} & k_{2} & k_{3} & k_{4} & k_{5} & k_{6}\\ \\
{k_{1}}^2 & {k_{2}}^2 & {k_{3}}^2 & {k_{4}}^2 & {k_{5}}^2 & {k_{6}}^2\\ \\

{e}^{(k_{1}L)} & {e}^{(k_{2}L)} & {e}^{(k_{3}L)} & {e}^{(k_{4}L)} & {e}^{(k_{5}L)} & {e}^{(k_{6}L)}\\ \\

k_{1}^{2}{e}^{(k_{1}L)} & k_{2}^{2}{e}^{(k_{2}L)} & k_{3}^{2}{e}^{(k_{3}L)} & k_{4}^{2}{e}^{(k_{4}L)} & k_{5}^{2}{e}^{(k_{5}L)} & k_{6}^{2}{e}^{(k_{6}L)}\\ \\
p_{1} & p_{2} & p_{3} & p_{4} & p_{5} & p_{6} \\ \\
\end{bmatrix}
$$

\noindent (e) Free-free beam :

$$[F(\omega)]=
\begin{bmatrix}
q_{1} & q_{2} & q_{3} & q_{4} & q_{5} & q_{6} \\ \\
r_{1} & r_{2} & r_{3} & r_{4} & r_{5} & r_{6} \\ \\
k_{1}^{3} & k_{2}^{3} & k_{3}^{3} & k_{4}^{3} & k_{5}^{3} & k_{6}^{3}\\ \\
t_{1} & t_{2} & t_{3} & t_{4} & t_{5} & t_{6} \\ \\
p_{1} & p_{2} & p_{3} & p_{4} & p_{5} & p_{6} \\ \\
k_{1}^{3}{e}^{(k_{1}L)} & k_{2}^{3}{e}^{(k_{2}L)} & k_{3}^{3}{e}^{(k_{3}L)} & k_{4}^{3}{e}^{(k_{4}L)} & k_{5}^{3}{e}^{(k_{5}L)} & k_{6}^{3}{e}^{(k_{6}L)}\\ \\
\end{bmatrix}
$$

\noindent Where,  \\

\noindent $t_{1}=(k_{1}^{3}-g^{2}{k_{1}}^{5}){e}^{(k_{1}L)} ,\quad  t_{2}=(k_{2}^{3}-g^{2}{k_{2}}^{5}){e}^{(k_{2}L)} ,\quad t_{3}=(k_{3}^{3}-g^{2}{k_{3}}^{5}){e}^{(k_{3}L)}$ \\ \\
$t_{4}=(k_{4}^{3}-g^{2}{k_{4}}^{5}){e}^{(k_{4}L)}  ,\quad t_{5}=(k_{5}^{3}-g^{2}{k_{5}}^{5}){e}^{(k_{5}L)} \quad t_{6}=(k_{6}^{3}-g^{2}{k_{6}}^{5}){e}^{(k_{6}L)}$ \\

\noindent $p_{1}=(k_{1}^{2}-g^{2}{k_{1}}^{4}){e}^{(k_{1}L)} ,\quad p_{2}=(k_{2}^{2}-g^{2}{k_{2}}^{4}){e}^{(k_{2}L)} ,\quad p_{3}=(k_{3}^{2}-g^{2}{k_{3}}^{4}){e}^{(k_{3}L)}$\\ \\
 $p_{4}=(k_{4}^{2}-g^{2}{k_{4}}^{4}){e}^{(k_{4}L)} ,\quad p_{5}=(k_{5}^{2}-g^{2}{k_{5}}^{4}){e}^{(k_{5}L)} , \quad p_{6}=(k_{6}^{2}-g^{2}{k_{6}}^{4}){e}^{(k_{6}L)}$ \\
 
\noindent $q_{1}=(k_{1}^{3}-g^{2}{k_{1}}^{5}) ,\quad  q_{2}=(k_{2}^{3}-g^{2}{k_{2}}^{5}) ,\quad q_{3}=(k_{3}^{3}-g^{2}{k_{3}}^{5})$ \\ \\
$q_{4}=(k_{4}^{3}-g^{2}{k_{4}}^{5})  ,\quad q_{5}=(k_{5}^{3}-g^{2}{k_{5}}^{5}) \quad q_{6}=(k_{6}^{3}-g^{2}{k_{6}}^{5})$ \\

\noindent $r_{1}=(k_{1}^{2}-g^{2}{k_{1}}^{4}) ,\quad r_{2}=(k_{2}^{2}-g^{2}{k_{2}}^{4}) ,\quad r_{3}=(k_{3}^{2}-g^{2}{k_{3}}^{4})$\\ \\
 $r_{4}=(k_{4}^{2}-g^{2}{k_{4}}^{4}) ,\quad r_{5}=(k_{5}^{2}-g^{2}{k_{5}}^{4}) , \quad r_{6}=(k_{6}^{2}-g^{2}{k_{6}}^{4})$ \\

\end{document}